\documentclass[twocolumn,showpacs,preprintnumbers,amsmath,amssymb]{revtex4}
\usepackage{graphicx}% Include figure files
\usepackage{dcolumn}% Align table columns on decimal point
\usepackage{bm}% bold math

\newcommand{\eigreim}[6]  %imput is the coordinates for picture
{\begin{picture}(#1,#2)(#3,#4)
\put(0,0){\line(1,0){#5}}
\put(0,0){\line(0,1){#5}}
\put(0,0){\line(-1,0){#5}}
\put(0,0){\line(0,-1){#5}}
\put(#6,0.0){\circle*{#6}}
\put(-#6,0.0){\circle*{#6}}
\put(0.0,#6){\circle*{#6}}
\put(0.0,-#6){\circle*{#6}}
\end{picture}
}

\newcommand{\eigreimout}[6]  %imput is the coordinates for picture
{\begin{picture}(#1,#2)(#3,#4)
\put(0,0){\line(1,0){#5}}
\put(0,0){\line(0,1){#5}}
\put(0,0){\line(-1,0){#5}}
\put(0,0){\line(0,-1){#5}}
\put(-#6,#6){\circle*{#6}}
\put(#6,#6){\circle*{#6}}
\put(-#6,-#6){\circle*{#6}}
\put(#6,-#6){\circle*{#6}}
\end{picture}
}

\newcommand{\eigrenegnegpospos}[6]  %imput is the coordinates for picture
{\begin{picture}(#1,#2)(#3,#4)
\put(0,0){\line(1,0){#5}}
\put(0,0){\line(0,1){#5}}
\put(0,0){\line(-1,0){#5}}
\put(0,0){\line(0,-1){#5}}
\put(-1.2,0.0){\circle*{#6}}
\put(-#5,0.0){\circle*{#6}}
\put(#5,0.0){\circle*{#6}}
\put(1.2,0.0){\circle*{#6}}
\end{picture}
}

\newcommand{\eigimnegnegpospos}[6]  %imput is the coordinates for picture
{\begin{picture}(#1,#2)(#3,#4)
\put(0,0){\line(1,0){#5}}
\put(0,0){\line(0,1){#5}}
\put(0,0){\line(-1,0){#5}}
\put(0,0){\line(0,-1){#5}}
\put(0.0,-1.2){\circle*{#6}}
\put(0.0,-#5){\circle*{#6}}
\put(0.0,#5){\circle*{#6}}
\put(0.0,1.2){\circle*{#6}}
\end{picture}
}

\newcommand{\eigrenegpos}[6]  %imput is the coordinates for picture
{\begin{picture}(#1,#2)(#3,#4)
\put(0,0){\line(1,0){#5}}
\put(0,0){\line(0,1){#5}}
\put(0,0){\line(-1,0){#5}}
\put(0,0){\line(0,-1){#5}}
\put(-3,-1){{\large $\times$}}
\put(0,-1){{\large$\times$}}
\end{picture}
}

\newcommand{\eigimnegpos}[6]  %imput is the coordinates for picture
{\begin{picture}(#1,#2)(#3,#4)
\put(0,0){\line(1,0){#5}}
\put(0,0){\line(0,1){#5}}
\put(0,0){\line(-1,0){#5}}
\put(0,0){\line(0,-1){#5}}
\put(-1.65,0.5){{\large $\times$}}
\put(-1.65,-2.5){{\large $\times$}}
\end{picture}
}

\newcommand{\eigzerorenegpos}[7]  %imput is the coordinates for picture
{\begin{picture}(#1,#2)(#3,#4)
\put(0,0){\line(1,0){#5}}
\put(0,0){\line(0,1){#5}}
\put(0,0){\line(-1,0){#5}}
\put(0,0){\line(0,-1){#5}}
\put(-1.65,-1){{\large $\times$}}
\put(-2,0.0){\circle*{#6}}
\put(2,0.0){\circle*{#6}}
\end{picture}
}

\newcommand{\eigzeroimnegpos}[7]  %imput is the coordinates for picture
{\begin{picture}(#1,#2)(#3,#4)
\put(0,0){\line(1,0){#5}}
\put(0,0){\line(0,1){#5}}
\put(0,0){\line(-1,0){#5}}
\put(0,0){\line(0,-1){#5}}
\put(-1.65,-1){{\large $\times$}}
\put(0.0,-2){\circle*{#6}}
\put(0.0,2){\circle*{#6}}
\end{picture}
}

\newcommand{\ofour}[7]  %imput is the coordinates for picture
{\begin{picture}(#1,#2)(#3,#4)
\put(0,0){\line(1,0){#5}}
\put(0,0){\line(0,1){#5}}
\put(0,0){\line(-1,0){#5}}
\put(0,0){\line(0,-1){#5}}
\put(-2.3,-1.5){{\LARGE $\times$}}%%%{\line(1,1){#7}}
\put(0.0,0.0){\circle{#6}}
\end{picture}
}

\begin{document}

%\preprint{APS/123-QED}

\title{Stability Analysis of the Lugiato-Lefever Model for Kerr Optical Frequency Combs. \\
       Part II: Case of Anomalous Dispersion}
\author{Irina Balakireva$^1$, Aur\'elien Coillet$^1$, Cyril Godey$^2$, and Yanne K. Chembo${^1}$}
\thanks{Corresponding author. E-mail: yanne.chembo@femto-st.fr}
\affiliation{$^1$FEMTO-ST Institute [CNRS UMR6174], Optics Department, \\
             16 Route de Gray, 25030 Besan\c con cedex, France. \\
             $^2$University of Franche-Comt\'e,  Department of Mathematics [CNRS UMR6623], \\
             16 Route de Gray, 25030 Besan\c con cedex, France.}
\date{\today}

\begin{abstract}
We present a stability analysis of the Lugiato-Lefever model for Kerr optical frequency combs in whispering gallery mode resonators pumped in the anomalous dispersion regime.
This article is the second part of a research work whose first part was devoted to the regime of normal dispersion, and was presented in ref.~\cite{Part_I}. 
The case of anomalous dispersion is indeed the most interesting from the theoretical point of view, because of the considerable variety of dynamical behaviors that can be observed.
From a technological point of view, it is also the most relevant because it corresponds to the regime where Kerr combs are predominantly generated, studied, and used for different applications.  
In this article, we analyze the connection between the spatial patterns and the bifurcation structure of the eigenvalues associated to the various equilibria of the system.
The bifurcation map evidences a considerable richness from a dynamical standpoint. We study in detail the emergence of 
super- and sub-critical Turing patterns in the system.
We determine the areas were bright isolated cavity solitons emerge, and we show that soliton molecules can emerge as well. Very complex temporal patterns can actually be observed in the system, where solitons (or soliton complexes) co-exist with or without mutual interactions. Our investigations also unveil the mechanism leading to the phenomenon of breathing solitons. Two routes to chaos in the system are identified, namely a route via the so called secondary combs, and another via soliton breathers. The Kerr combs corresponding to all these temporal patterns are analyzed in detail, and a discussion is led about the possibility to gain synthetic comprehension of the observed spectra out of the dynamical complexity of the system.
\end{abstract}

\maketitle

\section{Introduction}
\label{intro}

The study of Kerr optical frequency comb generation has been the focus of extensive research efforts in recent years.
These spectral grids are obtained through pumping a ultra-high $Q$ whispering-gallery mode (WGM) resonator with a narrow-linewidth continuous (CW) laser. 

Almost all previous research have been devoted to the investigation of Kerr comb generation in the anomalous group velocity dispersion (GVD) regime. In fact, it was thought for a long time that normal GVD Kerr comb generation was impossible, and later on, it has been shown to occur only under fairly exceptional circumstances (see for example refs.~\cite{Matsko_Normal,IEEE_PJ}).
This explain why the quasi-totality of the scientific literature on Kerr combs assumes a laser pump frequency in the anomalous GVD regime for the bulk material of the resonator.
The theoretical explanation of the difficulty to generate Kerr combs in the normal GVD regime was the purpose of
the first part of this research work, presented in ref.~\cite{Part_I}.

When the dispersion is anomalous, earlier studies on Kerr comb generation have shown that above a given threshold,
the long-lifetime photons originating from the pump interact nonlinearly with the medium and populate the neighboring cavity modes through four-wave mixing (FWM). The resulting permanent state features an all-to-all coupling amongst the excited modes, which can enable various dynamical outputs such as phase-locked (through Turing patterns or solitons), pulsating and even chaotic states.
In particular, the phase-locked states are expected to be useful for a wide spectrum of  applications~\cite{Review_Kerr_combs_Science,Nature_Ferdous,Nature_Selectable_freq,PRA_Scott2,PRL_NIST,Nature_Universal,Nature_Mid_IR,Jove}.

Synthetic studies where all these behaviors are associated to well identified regions of the parameter space are scarce. 
Most research articles so far have focused on specific phenomenologies (modulational instability, solitons, chaos, breathers, etc.) and our objective in this paper is to provide a larger viewpoint for the understanding of Kerr comb generation with anomalous GVD. 
The models used so far to investigate Kerr comb generation are either based on a modal expansion approach~\cite{Maleki_PRL_LowThres,YanneNanPRL,YanneNanPRA} or on a spatiotemporal formalism~\cite{GaetaOE,Matsko_OL_2,PRA_Yanne-Curtis,Coen} which is a variant of the Lugiato-Lefever equation (LLE)~\cite{LL}.
Both approaches can be shown to be equivalent under certain conditions~\cite{PRA_Yanne-Curtis}, but the latter appears to be more powerful to investigate the collective dynamics of a large number of modes. 

%%%%%%%%%%%%%%%%%%%%%%%%%%%%%%%%%%%%%%%%%%%%%%%%%%%%%%%%%%%%%%%%%%%%%%%%%%%%%%%%%%%%%%%%%%%%%%%%%%%%%%%%%%%%%%%%%%%%%%%%%%%%%%%%%%%%%%%%%%%%%%%%%%%%%%%%%%%
%%%%%%%%%%%%%%%%%%%%%%%%%%%%%%%%%%%%%%%%%%%%%%%%%%%%%%%%%%%%%%%%%%%%%%%%%%%%%%%%%%%%%%%%%%%%%%%%%%%%%%%%%%%%%%%%%%%%%%%%%%%%%%%%%%%%%%%%%%%%%%%%%%%%%%%%%%%
%%%%%%%%%%%%%%%%%%%%%%%%%%%%%%%%%%%%%%%%%%%%%%%%%%%%%%%%%%%%%%%%%%%%%%%%%%%%%%%%%%%%%%%%%%%%%%%%%%%%%%%%%%%%%%%%%%%%%%%%%%%%%%%%%%%%%%%%%%%%%%%%%%%%%%%%%%%
\begin{figure}
\begin{center}
\includegraphics[width=8cm]{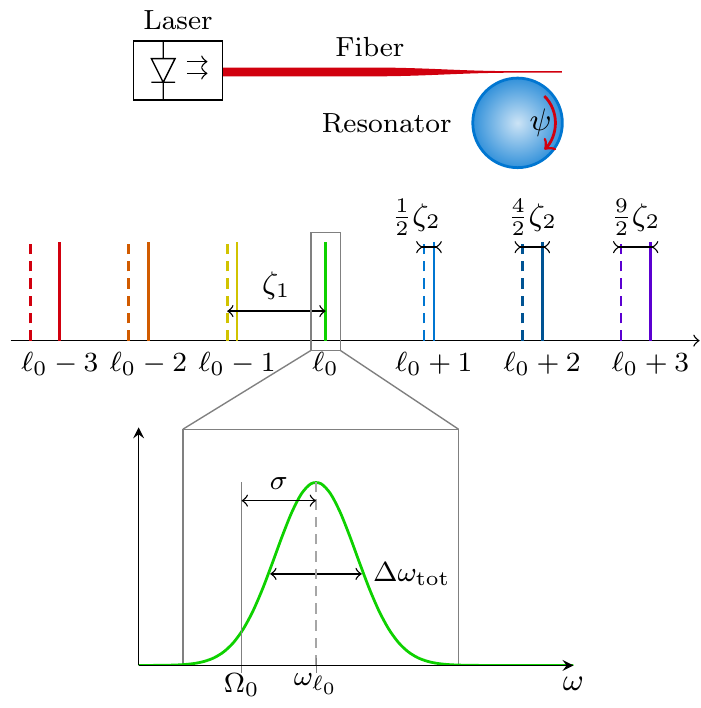}
\end{center}
\caption[Figure1]
{\label{Figure1} (Color online) 
(a) Schematical representation of a WGM resonator pumped with CW laser.
(b) Eigenmodes of the resonator. 
The real location of the modes with anomalous dispersion is represented in solid lines, while the dashed lines represent the location of the modes if the dispersion was null (perfect equidistance). 
Note that the anomalous dispersion pulls the modes rightwards (blueshift).
}
\end{figure}
%%%%%%%%%%%%%%%%%%%%%%%%%%%%%%%%%%%%%%%%%%%%%%%%%%%%%%%%%%%%%%%%%%%%%%%%%%%%%%%%%%%%%%%%%%%%%%%%%%%%%%%%%%%%%%%%%%%%%%%%%%%%%%%%%%%%%%%%%%%%%%%%%%%%%%%%%%%
%%%%%%%%%%%%%%%%%%%%%%%%%%%%%%%%%%%%%%%%%%%%%%%%%%%%%%%%%%%%%%%%%%%%%%%%%%%%%%%%%%%%%%%%%%%%%%%%%%%%%%%%%%%%%%%%%%%%%%%%%%%%%%%%%%%%%%%%%%%%%%%%%%%%%%%%%%%
%%%%%%%%%%%%%%%%%%%%%%%%%%%%%%%%%%%%%%%%%%%%%%%%%%%%%%%%%%%%%%%%%%%%%%%%%%%%%%%%%%%%%%%%%%%%%%%%%%%%%%%%%%%%%%%%%%%%%%%%%%%%%%%%%%%%%%%%%%%%%%%%%%%%%%%%%%%

In this article which is the second part of ref.~\cite{Part_I}, we study the bifurcation structure related to the different steady state solutions when the nonlinear WGM resonator is pumped in the anomalous GVD regime.
Our parameter space will be two-dimensional, and will involve the pump power and the laser detuning relatively to the cavity resonance. These parameters have been chosen because of their experimental relevance.

The plan of the article is the following.
In Sec.~\ref{model_eq_stab}, we briefly present the model, as well as the various equilibria and the related spatial bifurcations. Then, a synthetic bifurcation map is obtained and discussed in Sec.~\ref{bifurcationmap}. 
Turing patterns (rolls) arising from modulational instability (MI) are shown to emerge in the system, and they are investigated in Sec.~\ref{turingpatterns}.
In particular, their super- or sub-critical nature is analyzed in the time domain with respect to the detuning frequency of the pump laser. The MI gain is also determined analytically in our case and enables to understand the main properties of these patterns, such that the number of rolls, or their growth rate as the pump power of the laser is increased.
We study the emergence of bright cavity solitons and soliton molecules in Secs.~\ref{brightcavitysolitons} and 
Sec.~\ref{brightsolitonsmolecules} respectively. The connexion between these solitons and sub-critical Turing patterns is evidenced and enables to understand their main dynamical properties.
We also show that soliton molecules can coexist in the resonator, and generate very complex Kerr comb spectra. 
Breather solitons are studied in Sec.~\ref{Breathingsolitons}, and we show that the breathing period is necessarily of the order of magnitude of the photon lifetime.  
Chaotic behavior is investigated in Sec.~\ref{Chaos}, and we identify two main routes to chaos, either through the destabilization Turing patterns through  higher order bifurcations, of from the destabilization of solitonic structures.
Section~\ref{influencedispersion} focuses on the study of the effects of the magnitude of the dispersion parameter in
the dissipative structures that are formed in the cavity.
The last section of this article resumes our results.

%%%%%%%%%%%%%%%%%%%%%%%%%%%%%%%%%%%%%%%
\begin{table*}
\begin{center}
%\begin{ruledtabular}
\begin{tabular}{|c|c|c|c|c|}
\hline 
\multicolumn{5}{|c|} {\textbf{}}\\
\multicolumn{5}{|c|}{\textbf{Eigenvalues and reversible spatial bifurcations in the system}} \\
\multicolumn{5}{|c|}{\textbf{}}\\
\hline Denomination & Eigenvalues $(\lambda_{1,2};\lambda_{3,4})$ & Pictogram &  Bifurcation   & Location in Fig.~\ref{Figurebifnotatscale}     \\
\hline
                     &                 &      &     &   \\
Type~1      & $(\pm a; \pm b)$ & \eigrenegnegpospos{-2}{1}{1}{-0.8}{3}{1.5} &     &   \\
                     &                 &       &     &  \\
\hline
                     &                 &       &     &  \\
Type~2     & $(0;0)$ & \ofour{-2}{1}{1}{-0.8}{3}{3.7}{3.6}  & ${0}^4$    & ${\mathsf{a}}$ \\
                     &                 &      &     &   \\
\hline
                     &                 &       &     &  \\
Type~3    & $(\pm ia; \pm ib)$ & \eigimnegnegpospos{-2}{1}{1}{-0.7}{3}{1.5}  &     &  \\
                     &                 &       &     &  \\
\hline
                     &                 &       &     &  \\
Type~4     & $(\pm a; 0)$ & \eigzerorenegpos{-2}{1}{1}{-0.7}{3}{1.5}{1.5} & ${0}^2$ & ${\mathsf{B}}_1$  \\
                     &                 &       &     &  \\
\hline
                     &                 &       &     &  \\
Type~5      & $(0; \pm ib)$ & \eigzeroimnegpos{-2}{1}{1}{-0.7}{3}{1.5}{1.5}  & ${0}^2 (i \omega)$  &  ${\mathsf{B}}_2$, ${\mathsf{b}}$, ${\mathsf{C}}_1$, ${\mathsf{c}}$, ${\mathsf{C}}_2$  \\
                     &                 &       &     &  \\
\hline
                     &                 &      &     &   \\
Type~6     & $(\pm a; \pm ib)$ & \eigreim{-2}{1}{1}{-0.7}{3}{1.5}  &     &  \\
                     &                 &      &     &   \\
\hline
                     &                 &       &     &  \\
Type~7    & $(\pm ia; \pm ia)$ & \eigimnegpos{-2}{1}{1}{-0.7}{3}{1.5}  & $(i \omega)^2$ & ${\mathsf{A}}_1$ , ${\mathsf{A}}_2$ \\
                     &                 &      &     &   \\
\hline
                     &                 &       &     &  \\
Type~8   & $(\pm a; \pm a)$ & \eigrenegpos{-2}{1}{1}{-0.7}{3}{4}  &     &   \\
                     &                 &       &     &  \\
\hline
                     &                 &      &     &   \\
Type~9     & $(a \pm ib; c \pm id)$ & \eigreimout{-2}{1}{1}{-0.7}{3}{1.5}  &     &  \\
                     &                 &      &     &   \\
\hline
\end{tabular}
%\end{ruledtabular}
\end{center}
\caption{Nomenclature and pictograms for the various sets of eigenvalues.
A set of four eigenvalues is attached to each equilibrium, and some classified bifurcations are attached to certain configurations of eigenvalues.
A dot stands for a single eigenvalue, the cross stands for a set of two degenerated eigenvalues (double non semi-simple eigenvalue), and a circled cross corresponds to a set of four degenerated eigenvalues (quadruple eigenvalue with a $4 \times 4$ Jordan bloc).}
\label{NomenclatureBifurcations}
\end{table*}
%%%%%%%%%%%%%%%%%%%%%%%%%%%%%%%%%%%%%%%

\section{Model, equilibria and spatial bifurcations}
\label{model_eq_stab}

As in the first part of this study, the system is a WGM disk pumped by a CW pump laser.
as displayed in Fig.~\ref{Figure1}. For the sake of the self-consistency of this article, we very briefly recall the main features of the model in the next subsection.

\subsection{Model and equilibria}
\label{model_eq}

The eigenvalues of the WGM cavity are Taylor-expanded as $\omega_\ell = \omega_{\ell_0} + \zeta_1 (\ell - \ell_0) + \frac{1}{2}\zeta_2 (\ell - \ell_0)^2 $, where $\ell$ is the integer eigennumber that unambigously labels the modes belonging to the fundamental (or toric) family of WGMs, while $\zeta_1$ and  $\zeta_2$ are respectively the free-spectral range (FSR) of the WGM resonator and the second-order dispersion coefficient. The FSR obeys $\zeta_1 = c/ a n_0 = 2 \pi /T$, with $c$ being the velocity of light in vacuum, $a$ the main radius of the disk, $n_0$ is the refraction index of the disk for a frequency equal to the resonance $\omega_{\ell_0} \simeq  \zeta_1 \, \ell_0 $, and $T$ being the round-trip time.

In its normalized form, the total intracavity field obeys the following partial differential equation when the modal linewidths are quasi-degenerated~\cite{YanneNanPRA} 
\begin{eqnarray}
\frac{\partial \psi}{\partial \tau}  =  - (1 +i\alpha) \psi + i |\psi|^2 \psi - i \frac{\beta}{2} \frac{\partial^2 \psi}{ \partial \theta^2 } +F \, ,
\label{final_eq_dimensionless}
\end{eqnarray}
where  $\psi (\theta ,\tau)$ is the complex envelope of the total intra-cavity
field, $\theta \in [-\pi,\pi]$ is the azimuthal angle along the circumference, and $\tau=  \Delta \omega_{\rm tot} t /2$ is the dimensionless time, with $\Delta \omega_{\rm tot}$ being the loaded (total) modal linewidth. 
This LLE has periodic boundary conditions, and $\psi$ represents the intra-cavity fields dynamics in the moving frame~\cite{IEEE_PJ}.
The parameters of this equation are: the frequency detuning $\alpha = - {2 (\Omega_0 -\omega_{\ell_0})}/{\Delta \omega_{\rm tot}}$ where $\Omega_0$ and $\omega_{\ell_0}$ are respectively the angular frequencies of the pumping laser and the cold-cavity resonance; the overall dispersion parameter $\beta = - {2 \zeta_2}/{\Delta \omega_{\rm tot}}$ (positive for normal GVD and negative for anomalous GVD); and finally the pump field intensity 
$F = [{8 g_0 \Delta \omega_{\rm ext}}/{\Delta \omega_{\rm tot}^3}]^{1/2} \, [{P}/{\hbar \Omega_0}]^{1/2}$
where $P$ is the intensity (in W) of the laser pump at the input of the
resonator, $g_0= n_2 c \hbar \Omega_0^2/n_0^2 V_0$ is the nonlinear gain, 
$n_0$ and $n_2$ are respectively the linear and nonlinear refraction indices of the bulk material, $V_0$ is the effective volume of the pumped mode, and $\Delta \omega_{\rm ext}$ being the extrinsic (coupling) modal linewidth~\cite{JSTQE}.
It is important to recall that $\psi$ represents the intracavity field in a moving frame that is circumferentially rotating at frequency $\zeta_1$. 
The order of magnitude for these various parameters are given and commented in Part~I~\cite{Part_I}.  
As demonstrated in ref.~\cite{PRA_Yanne-Curtis}, the intracavity field can be expanded as
\begin{eqnarray}
\psi (\theta ,\tau)  = \sqrt{\frac{2 g_0 }{\Delta \omega_{\rm tot}}}
                      \sum_{\ell} {\cal A}_\ell^* (\tau)  e^{\left[i (\ell - \ell_0) \theta + i\frac{1}{2}\beta (\ell - \ell_0)^2 \tau  \right]} \, ,
\label{psi_expansion}
\end{eqnarray}
where ${\cal A}_\ell$ corresponds to the normalized modal fields introduced in refs.~\cite{YanneNanPRL,YanneNanPRA}.
\\

%%%%%%%%%%%%%%%%%%%%%%%%%%%%%%%%%%%%%%%%%%%%%%%%%%%%%%%%%%%%%%%%%%%%%%%%%%%%%%%%%%%%%%%%%%%%%%%%%%%%%%%%%%%%%%%%%%%%%%%%%%%%%%%%%%%%%%%%%%%%%%%%%%%%%%%%%%%
%%%%%%%%%%%%%%%%%%%%%%%%%%%%%%%%%%%%%%%%%%%%%%%%%%%%%%%%%%%%%%%%%%%%%%%%%%%%%%%%%%%%%%%%%%%%%%%%%%%%%%%%%%%%%%%%%%%%%%%%%%%%%%%%%%%%%%%%%%%%%%%%%%%%%%%%%%%
%%%%%%%%%%%%%%%%%%%%%%%%%%%%%%%%%%%%%%%%%%%%%%%%%%%%%%%%%%%%%%%%%%%%%%%%%%%%%%%%%%%%%%%%%%%%%%%%%%%%%%%%%%%%%%%%%%%%%%%%%%%%%%%%%%%%%%%%%%%%%%%%%%%%%%%%%%%
\begin{figure*}
\begin{center}
\includegraphics[width=17cm]{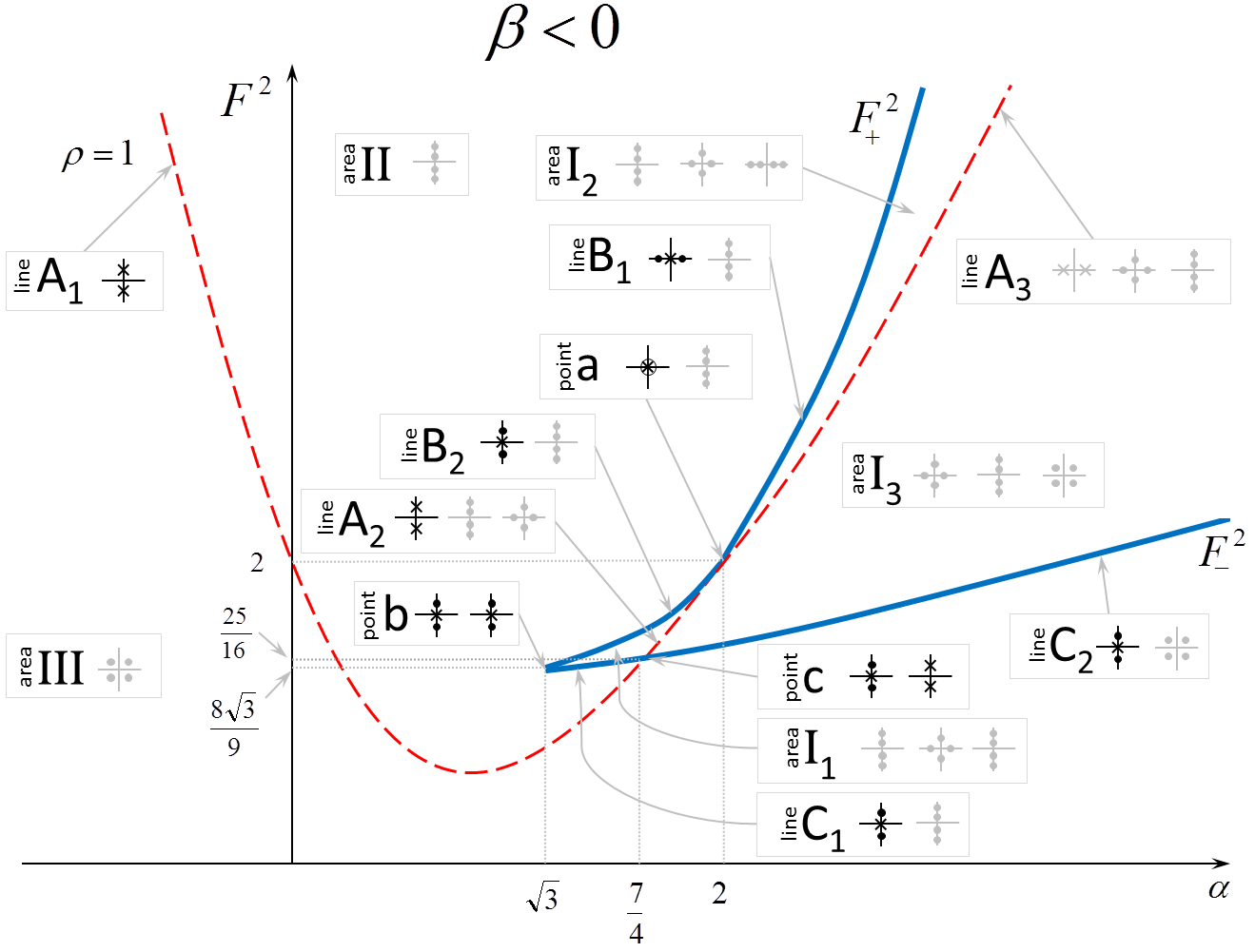}
\end{center}
\caption[Figurebif]
{\label{Figurebifnotatscale} (Color online) Bifurcation diagram (not at scale).
The areas are labelled using Roman numbers (I, II, and III).
The lines are labelled using capital letters, and line ${\mathsf{A}}$ stands for the limit $\rho = 1$ (dashed line in the figure), ${\mathsf{B}}$ stands for $F^2_+ (\alpha)$, and ${\mathsf{C}}$ standing for $F^2_- (\alpha)$.
The points are labelled using low-case letters (${\mathsf{a}}$ and ${\mathsf{b}}$). 
Both areas and lines can be divided in sub-domains (I$_1$, I$_2$, ${\mathsf{A}}_1$, ${\mathsf{B}}_1$, etc.)
Note that the system has three equilibria (three quadruplets of eigenvalues) in the area I, one equilibrium outside (one quadruplet), and two on the boundaries (two quadruplets). 
Black eigenvalue pictograms denote a bifurcation according to Table~\ref{NomenclatureBifurcations}, while grey pictograms do not.}
\end{figure*}
%%%%%%%%%%%%%%%%%%%%%%%%%%%%%%%%%%%%%%%%%%%%%%%%%%%%%%%%%%%%%%%%%%%%%%%%%%%%%%%%%%%%%%%%%%%%%%%%%%%%%%%%%%%%%%%%%%%%%%%%%%%%%%%%%%%%%%%%%%%%%%%%%%%%%%%%%%%
%%%%%%%%%%%%%%%%%%%%%%%%%%%%%%%%%%%%%%%%%%%%%%%%%%%%%%%%%%%%%%%%%%%%%%%%%%%%%%%%%%%%%%%%%%%%%%%%%%%%%%%%%%%%%%%%%%%%%%%%%%%%%%%%%%%%%%%%%%%%%%%%%%%%%%%%%%%
%%%%%%%%%%%%%%%%%%%%%%%%%%%%%%%%%%%%%%%%%%%%%%%%%%%%%%%%%%%%%%%%%%%%%%%%%%%%%%%%%%%%%%%%%%%%%%%%%%%%%%%%%%%%%%%%%%%%%%%%%%%%%%%%%%%%%%%%%%%%%%%%%%%%%%%%%%%

The  equilibria $\psi_{\rm e}$ of Eq.~(\ref{final_eq_dimensionless}) obey
\begin{eqnarray}
F^2 = [1+ (\rho^2 - \alpha )^2] \rho^2   \equiv G (\alpha,\rho)  \, .
\label{def_F2}
\end{eqnarray}
 with $\rho = |\psi_{\rm e}|^2$. This cubic equation in $\rho$ only has one solution when $\alpha < \sqrt{3}$ (always stable). When $\alpha> \sqrt{3}$,  there are three equilibria $\rho_1 \leq \rho_2 \leq \rho_3$ when $F^2 \in [F^2_-(\alpha), F^2_+(\alpha)]$, with 
$F^2_\pm (\alpha) = G[\alpha, \rho_\pm (\alpha)]$ and $\rho_\pm (\alpha) = [2 \alpha \pm \sqrt{\alpha^2 - 3}]/{3}$.
The intermediate solutions is always temporally unstable while the extremal solutions are always stable (hysteresis).

\subsection{Spatial bifurcations}
\label{spatialbifurcations}

The spatial bifurcation study in the anomalous GVD case is more complex than in the normal GVD regime because several solutions of interest depend on the boundary condition. Disregarding the finiteness of the $\theta$-domain is therefore not a correct assumption most of the time. 
However, since both the modal expansion and the spatiotemporal models already gave a solid theoretical understanding 
of the effects induced by the periodic boundary conditions, we can combine these previous knowledge with the fixed-points eigenvalue analysis in order to gain detailed understanding of the various solutions of the system.

As for the normal GVD case, we set $\partial_\tau \psi \equiv 0$  and introduce the intermediate variables 
$\phi_{r,i} = \partial_\theta \psi_{r,i}$ in order to rewrite the LLE under the form
\begin{eqnarray}
\frac{\partial \psi_r}{\partial \theta} &=& \phi_r \\
\frac{\partial \phi_r}{\partial \theta} &=& \frac{2}{\beta} ( \psi^3_r + \psi^2_i \psi_r - \alpha \psi_r - \psi_i)\\
\frac{\partial \psi_i}{\partial \theta} &=& \phi_i \\
\frac{\partial \phi_i}{\partial \theta} &=& \frac{2}{\beta} ( \psi^2_r \psi_i + \psi^3_i - \alpha \psi_i + \psi_r - F) \, ,
\label{4D_flow_eqs_stab_spatial}
\end{eqnarray}
where $\psi = \psi_r + i \psi_i$, with $\psi_r$ and $\psi_i$ being the real and imaginary parts of $\psi$, respectively.

%%%%%%%%%%%%%%%%%%%%%%%%%%%%%%%%%%%%%%%%%%%%%%%%%%%%%%%%%%%%%%%%%%%%%%%%%%%%%%%%%%%%%%%%%%%%%%%%%%%%%%%%%%%%%%%%%%%%%%%%%%%%%%%%%%%%%%%%%%%%%%%%%%%%%%%%%%%
%%%%%%%%%%%%%%%%%%%%%%%%%%%%%%%%%%%%%%%%%%%%%%%%%%%%%%%%%%%%%%%%%%%%%%%%%%%%%%%%%%%%%%%%%%%%%%%%%%%%%%%%%%%%%%%%%%%%%%%%%%%%%%%%%%%%%%%%%%%%%%%%%%%%%%%%%%%
%%%%%%%%%%%%%%%%%%%%%%%%%%%%%%%%%%%%%%%%%%%%%%%%%%%%%%%%%%%%%%%%%%%%%%%%%%%%%%%%%%%%%%%%%%%%%%%%%%%%%%%%%%%%%%%%%%%%%%%%%%%%%%%%%%%%%%%%%%%%%%%%%%%%%%%%%%%
\begin{figure}
\begin{center}
\includegraphics[width=4.6cm]{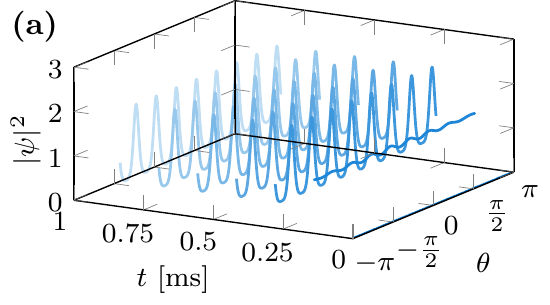}
\includegraphics[width=3.4cm]{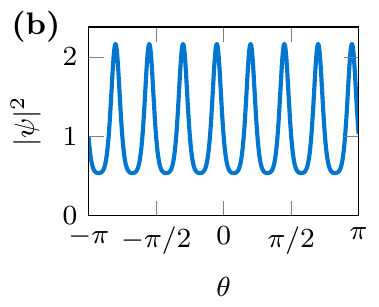}
\includegraphics[width=3.4cm]{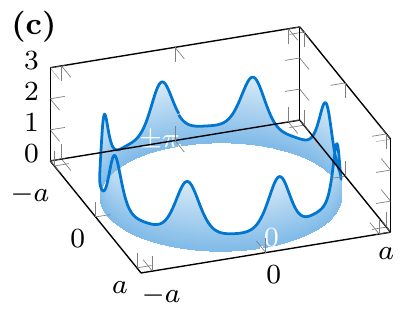}
\includegraphics[width=4.6cm]{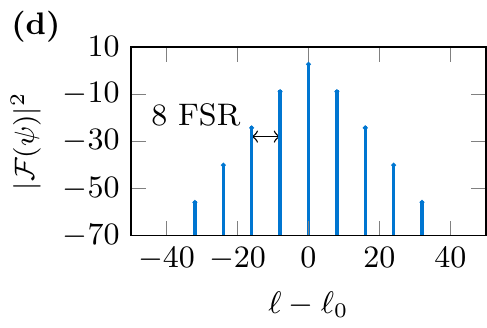}
\end{center}
\caption[Supercritical_Turing]
{\label{Supercritical_Turing} (Color online) Supercritical Turing patterns (so-called \textit{soft} excitation).
The parameters are $\alpha =1$, $\beta= -0.04$, and $\rho = 1.2$ 
[the pumping power $F^2$ can be directly calculated using Eq.~(\ref{def_F2})].
(a) Transient dynamics.
(b) Final pattern in the azimuthal direction.
(c) 3D representation.
(d) Corresponding Kerr comb.
 }
\end{figure}
%%%%%%%%%%%%%%%%%%%%%%%%%%%%%%%%%%%%%%%%%%%%%%%%%%%%%%%%%%%%%%%%%%%%%%%%%%%%%%%%%%%%%%%%%%%%%%%%%%%%%%%%%%%%%%%%%%%%%%%%%%%%%%%%%%%%%%%%%%%%%%%%%%%%%%%%%%%
%%%%%%%%%%%%%%%%%%%%%%%%%%%%%%%%%%%%%%%%%%%%%%%%%%%%%%%%%%%%%%%%%%%%%%%%%%%%%%%%%%%%%%%%%%%%%%%%%%%%%%%%%%%%%%%%%%%%%%%%%%%%%%%%%%%%%%%%%%%%%%%%%%%%%%%%%%%
%%%%%%%%%%%%%%%%%%%%%%%%%%%%%%%%%%%%%%%%%%%%%%%%%%%%%%%%%%%%%%%%%%%%%%%%%%%%%%%%%%%%%%%%%%%%%%%%%%%%%%%%%%%%%%%%%%%%%%%%%%%%%%%%%%%%%%%%%%%%%%%%%%%%%%%%%%%

The Jacobian matrix around an equilibrium $\psi_{\rm e} = \psi_{{\rm e},r} + i \psi_{{\rm e},i}$ is
 \begin{eqnarray}
\mathbf{J} =
            \begin{bmatrix}
                   0 & 1 & 0 & 0 \\
\frac{2}{\beta} ( 3 \psi_{{\rm e},r}^2\!+\! \psi_{{\rm e},i}^2 \!-\! \alpha ) &0& \frac{2}{\beta} (2 \psi_{
{\rm e},r} \psi_{{\rm e},i}\!-\! 1) & 0\\
0&0&0&1 \\
\frac{2}{\beta} (2 \psi_{{\rm e},r} \psi_{{\rm e},i}\!+\!1) &0&\frac{2}{\beta}(\psi_{{\rm e},r}^2\!\!+\! 3 \psi_{{\rm e},i}^2 \!-\! \alpha) & 0
            \end{bmatrix}
\end{eqnarray}
and the corresponding eigenvalues obey
\begin{eqnarray}
\lambda^4 \!+ \!\frac{4}{|\beta|}(2 \rho\! - \! \alpha) \lambda^2 \!+ \!\frac{4}{|\beta|^2} (3 \rho^2 \! - \!4 \alpha \rho \!+\! \alpha^2 \!+ \!1) =0 \, .
\label{polynom}
\end{eqnarray} 
Note that since $\beta$ is negative here (anomalous GVD), it has been rewritten as $-|\beta|$
in order to facilitate the comparison with the normal dispersion case studied in Part~I~\cite{Part_I}.

For being of fourth polynomial order, Eq.~(\ref{polynom}) will always yield a quadruplet of eigenvalues for each solution.
Hence, in $\alpha$--$F^2$ plane, there will be one quadruplet in the single-equilibrium area, three between the lines 
$F_\pm(\alpha)^2$, and two quadruplets onto these boundary lines.
We analyze below the nature (real, pure imaginary or complex) of the eigenvalues as a function of the sign of the discriminant $\Delta = 16 (\rho^2 - 1)$ of Eq.~(\ref{polynom}).

\subsubsection{First case: $\rho>1$}

The solutions of the paired solutions of Eq.~(\ref{polynom}) are  
\begin{eqnarray}
\lambda^2 \!=\! - \frac{2}{|\beta|} [2 \rho \!-\! \alpha \! \pm \! \sqrt{\rho^2 \!-\!1}]   \, .
\label{sol_discr_pos}
\end{eqnarray}
The product of these paired solutions is  $3 \rho^2 \! -\! 4 \alpha \rho \! +\! \alpha^2 \!+ \!1 $, which corresponds to $\partial G /\partial \rho$ (see Eq.~(\ref{def_F2})).

Three sub-cases have to be considered depending on the sign of $\partial G /\partial \rho$:
\begin{itemize}
\item If $\partial G /\partial \rho >0$, 
 the eigenvalues can be written
  as   $(\lambda_{1,2};\lambda_{3,4})=(\pm a; \pm b)$ if $2 \rho - \alpha <0$  (eigenvalues of Type~1); 
  as   $(\lambda_{1,2};\lambda_{3,4})=(0; 0)$ if $2 \rho - \alpha =0$ (Type~2), 
and as $(\lambda_{1,2};\lambda_{3,4})=(\pm ia; \pm ib)$ if $2 \rho - \alpha >0$ (Type~3).
\item If $\partial G /\partial \rho=0$, 
the eigenvalues can be written as 
      $(\lambda_{1,2};\lambda_{3,4})=(\pm a; 0)$ if $2 \rho - \alpha <0$ (Type~4); 
and as $(\lambda_{1,2};\lambda_{3,4})=(0; \pm ib)$ if $2 \rho - \alpha >0$ (Type~5). 
The eigenvalues degenerate here to Type~2 when $2 \rho - \alpha =0$.
\item If $\partial G /\partial \rho <0$, 
the eigenvalues have the form
      $(\lambda_{1,2};\lambda_{3,4})=(\pm a; \pm ib)$ (eigenvalue of Type~6).
\end{itemize}

\subsubsection{Second case:  $\rho=1$}

In this case, the solutions simply obey (double-root)
\begin{eqnarray}
\lambda^2 \!=\! - \frac{2}{|\beta|} [2  \!-\! \alpha ]   \, .
\label{sol_discr_null}
\end{eqnarray}
Hence, the eigenvalues can be written 
as $(\lambda_{1,2};\lambda_{3,4})=(\pm ia; \pm ia)$ when $\alpha <2$ (Type~7);
as $(\lambda_{1,2};\lambda_{3,4})=(0;0)$ when $\alpha =2$ (Type~2); 
and as $(\lambda_{1,2};\lambda_{3,4})=(\pm a; \pm a)$ when $\alpha >2$ (Type~8).

\subsubsection{Third case:  $\rho<1$}

Here, the eigenvalues are complex:
\begin{eqnarray}
\lambda^2 \!=\! - \frac{2}{|\beta|} [2 \rho \!-\! \alpha \! \pm \! i\sqrt{1 \!-\!\rho^2}]   \, .
\label{sol_discr_neg}
\end{eqnarray}
and they have the explicit form
$(\lambda_{1,2};\lambda_{3,4})=( a \pm ib ; c \pm id)$, here corresponding to Type~9.

%%%%%%%%%%%%%%%%%%%%%%%%%%%%%%%%%%%%%%%%%%%%%%%%%%%%%%%%%%%%%%%%%%%%%%%%%%%%%%%%%%%%%%%%%%%%%%%%%%%%%%%%%%%%%%%%%%%%%%%%%%%%%%%%%%%%%%%%%%%%%%%%%%%%%%%%%%%
%%%%%%%%%%%%%%%%%%%%%%%%%%%%%%%%%%%%%%%%%%%%%%%%%%%%%%%%%%%%%%%%%%%%%%%%%%%%%%%%%%%%%%%%%%%%%%%%%%%%%%%%%%%%%%%%%%%%%%%%%%%%%%%%%%%%%%%%%%%%%%%%%%%%%%%%%%%
%%%%%%%%%%%%%%%%%%%%%%%%%%%%%%%%%%%%%%%%%%%%%%%%%%%%%%%%%%%%%%%%%%%%%%%%%%%%%%%%%%%%%%%%%%%%%%%%%%%%%%%%%%%%%%%%%%%%%%%%%%%%%%%%%%%%%%%%%%%%%%%%%%%%%%%%%%%
\begin{figure}
\begin{center}
\includegraphics[width=4.6cm]{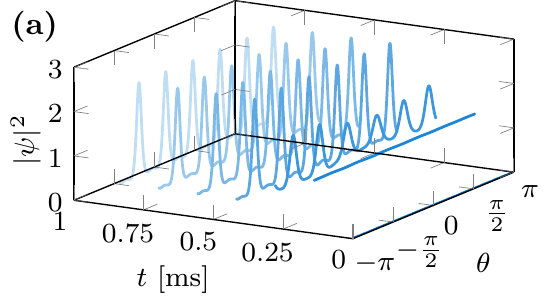}
\includegraphics[width=3.4cm]{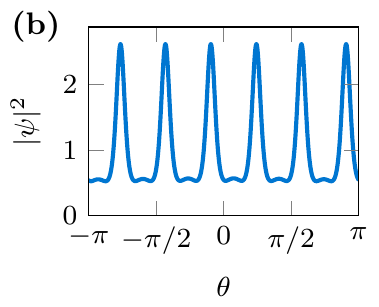}
\includegraphics[width=3.4cm]{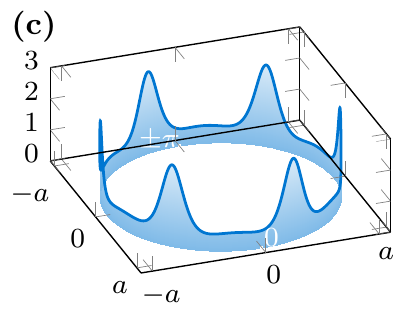}
\includegraphics[width=4.6cm]{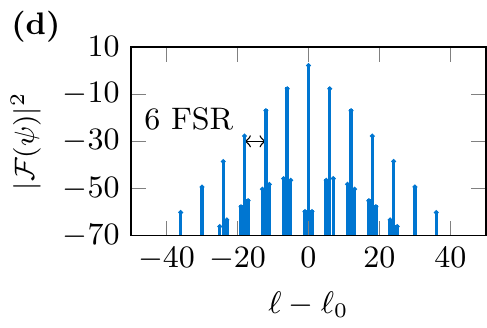}
\end{center}
\caption[Subcritical_Turing]
{\label{Subcritical_Turing} (Color online) Subcritical Turing patterns (so-called \textit{hard} excitation).
The parameters are $\alpha =1.5$, $\beta= -0.04$, and $\rho = 1.2$
[the pumping power $F^2$ can be directly calculated using Eq.~(\ref{def_F2})].
(a) Transient dynamics.
(b) Final pattern in the azimuthal direction.
(c) 3D representation.
(d) Corresponding Kerr comb.
 }
\end{figure}
%%%%%%%%%%%%%%%%%%%%%%%%%%%%%%%%%%%%%%%%%%%%%%%%%%%%%%%%%%%%%%%%%%%%%%%%%%%%%%%%%%%%%%%%%%%%%%%%%%%%%%%%%%%%%%%%%%%%%%%%%%%%%%%%%%%%%%%%%%%%%%%%%%%%%%%%%%%
%%%%%%%%%%%%%%%%%%%%%%%%%%%%%%%%%%%%%%%%%%%%%%%%%%%%%%%%%%%%%%%%%%%%%%%%%%%%%%%%%%%%%%%%%%%%%%%%%%%%%%%%%%%%%%%%%%%%%%%%%%%%%%%%%%%%%%%%%%%%%%%%%%%%%%%%%%%
%%%%%%%%%%%%%%%%%%%%%%%%%%%%%%%%%%%%%%%%%%%%%%%%%%%%%%%%%%%%%%%%%%%%%%%%%%%%%%%%%%%%%%%%%%%%%%%%%%%%%%%%%%%%%%%%%%%%%%%%%%%%%%%%%%%%%%%%%%%%%%%%%%%%%%%%%%%

\section{Bifurcation map}
\label{bifurcationmap}

The classification of the eigenvalues of the various fixed points in the $\alpha$--$F^2$ plane performed in the preceding section enables to obtain a stability map which is displayed in Fig.~\ref{Figurebifnotatscale}.
The various areas of this map are similar to the ones we have obtained in the normal GVD regime~\cite{Part_I}, where we had also explained that the $\theta \rightarrow -\theta$ symmetry in our system imposes that the relevant bifurcations  should necessarily be  \textit{reversible}
(see ref.~\cite{Mariana_Iooss}, Chapter~4). 
The following four bifurcations can emerge in Fig.~\ref{Figurebifnotatscale}:

\begin{itemize}

\item {${0}^2$ bifurcation}: 
also referred to as Takens-Bogdanov bifurcation, it arises when a quadruplet of eigenvalues
is of Type~4. This bifurcation occurs here along  the line ${\mathsf{B}}_1$.

\item {${0}^2 (i \omega)$ bifurcation}:
it corresponds to a quadruplet of Type~5, and it is present in our system in the lines 
${\mathsf{B}}_2$, ${\mathsf{C}}_1$
and ${\mathsf{C}}_2$ .

\item {$(i \omega)^2$ bifurcation}:
also known as the ``$1:1$ resonance'' or the ``Hamiltonian Hopf'' bifurcation, it emerges 
when a quadruplet of eigenvalues is of Type~7.
In Fig.~\ref{Figurebifnotatscale}, it corresponds to
the lines  ${\mathsf{A}}_1$ and  ${\mathsf{A}}_2$.

\item {${0}^4$ bifurcation}:
this co-dimension~2 bifurcation arises when a quadruplet of eigenvalues degenerates to the origin (Type~2).
This situation is only witnessed at the point ${ \rm a} \equiv (2,2)$.

\end{itemize}

%%%%%%%%%%%%%%%%%%%%%%%%%%%%%%%%%%%%%%%%%%%%%%%%%%%%%%%%%%%%%%%%%%%%%%%%%%%%%%%%%%%%%%%%%%%%%%%%%%%%%%%%%%%%%%%%%%%%%%%%%%%%%%%%%%%%%%%%%%%%%%%%%%%%%%%%%%%
%%%%%%%%%%%%%%%%%%%%%%%%%%%%%%%%%%%%%%%%%%%%%%%%%%%%%%%%%%%%%%%%%%%%%%%%%%%%%%%%%%%%%%%%%%%%%%%%%%%%%%%%%%%%%%%%%%%%%%%%%%%%%%%%%%%%%%%%%%%%%%%%%%%%%%%%%%%
%%%%%%%%%%%%%%%%%%%%%%%%%%%%%%%%%%%%%%%%%%%%%%%%%%%%%%%%%%%%%%%%%%%%%%%%%%%%%%%%%%%%%%%%%%%%%%%%%%%%%%%%%%%%%%%%%%%%%%%%%%%%%%%%%%%%%%%%%%%%%%%%%%%%%%%%%%%
\begin{figure}
\begin{center}
\includegraphics[width=3.5cm]{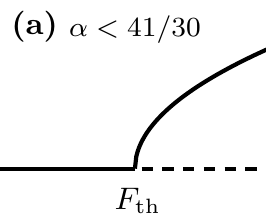}
~~
\includegraphics[width=3.5cm]{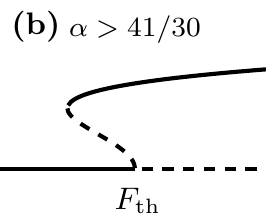}
\end{center}
\caption[super_sub]
{\label{super_sub} (Color online) Diagram showing the difference between supercritical ($\alpha < 41/30$) and subcritical  ($\alpha > 41/30$) pitchfork bifurcation towards Turing patterns. The continuous line denotes a stable amplitude while the dashed line stands for a unstable steady state. 
(a) In the supercritical case, no Kerr comb is possible below the threshold pump power $F_{\rm th}^2$.
Above this threshold, the sidemodes of the Kerr comb can continuously grow from infinitesimally small to significantly large as the pump is increased.
The stable equilibrium and the Turing pattern are never simultaneously stable. 
This excitation mode is sometimes referred to as \textit{soft}.
(b) In the subcritical case, Kerr comb is possible in a small interval \emph{below} the threshold pump power $F_{\rm th}^2$. Moreover, in the small interval below $F_{\rm th}^2$ where the comb can be excited, a stable Turing pattern does coexist with the equilibrium solution (bistability).  As a consequence, the transition from the flat state to the Turing pattern is abrupt, and the sidemodes of the comb can not be infinitesimally small. 
This excitation mode is sometimes referred to as \textit{hard}.
 }
\end{figure}
%%%%%%%%%%%%%%%%%%%%%%%%%%%%%%%%%%%%%%%%%%%%%%%%%%%%%%%%%%%%%%%%%%%%%%%%%%%%%%%%%%%%%%%%%%%%%%%%%%%%%%%%%%%%%%%%%%%%%%%%%%%%%%%%%%%%%%%%%%%%%%%%%%%%%%%%%%%
%%%%%%%%%%%%%%%%%%%%%%%%%%%%%%%%%%%%%%%%%%%%%%%%%%%%%%%%%%%%%%%%%%%%%%%%%%%%%%%%%%%%%%%%%%%%%%%%%%%%%%%%%%%%%%%%%%%%%%%%%%%%%%%%%%%%%%%%%%%%%%%%%%%%%%%%%%%
%%%%%%%%%%%%%%%%%%%%%%%%%%%%%%%%%%%%%%%%%%%%%%%%%%%%%%%%%%%%%%%%%%%%%%%%%%%%%%%%%%%%%%%%%%%%%%%%%%%%%%%%%%%%%%%%%%%%%%%%%%%%%%%%%%%%%%%%%%%%%%%%%%%%%%%%%%%

%%%%%%%%%%%%%%%%%%%%%%%%%%%%%%%%%%%%%%%%%%%%%%%%%%%%%%%%%%%%%%%%%%%%%%%%%%%%%%%%%%%%%%%%%%%%%%%%%%%%%%%%%%%%%%%%%%%%%%%%%%%%%%%%%%%%%%%%%%%%%%%%%%%%%%%%%%%
%%%%%%%%%%%%%%%%%%%%%%%%%%%%%%%%%%%%%%%%%%%%%%%%%%%%%%%%%%%%%%%%%%%%%%%%%%%%%%%%%%%%%%%%%%%%%%%%%%%%%%%%%%%%%%%%%%%%%%%%%%%%%%%%%%%%%%%%%%%%%%%%%%%%%%%%%%%
%%%%%%%%%%%%%%%%%%%%%%%%%%%%%%%%%%%%%%%%%%%%%%%%%%%%%%%%%%%%%%%%%%%%%%%%%%%%%%%%%%%%%%%%%%%%%%%%%%%%%%%%%%%%%%%%%%%%%%%%%%%%%%%%%%%%%%%%%%%%%%%%%%%%%%%%%%%
\begin{figure}
\begin{center}
\includegraphics[width=7.5cm]{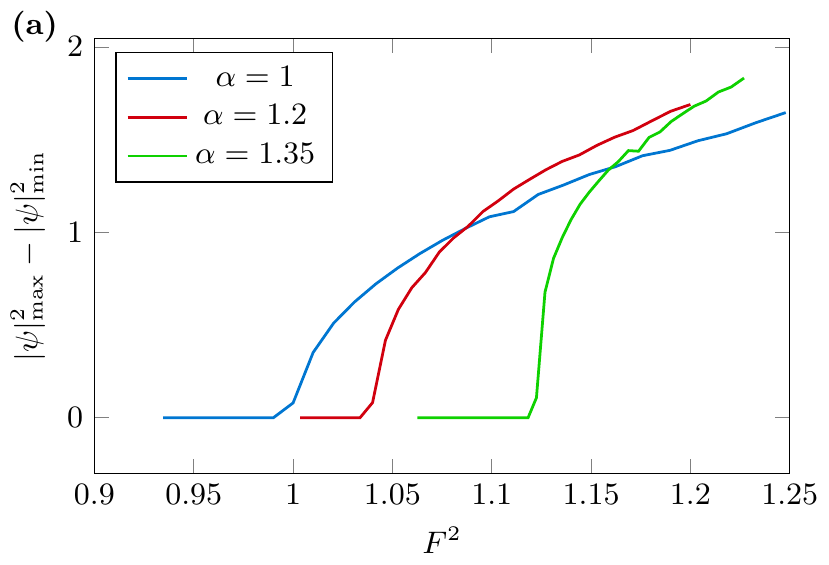}
\includegraphics[width=7cm]{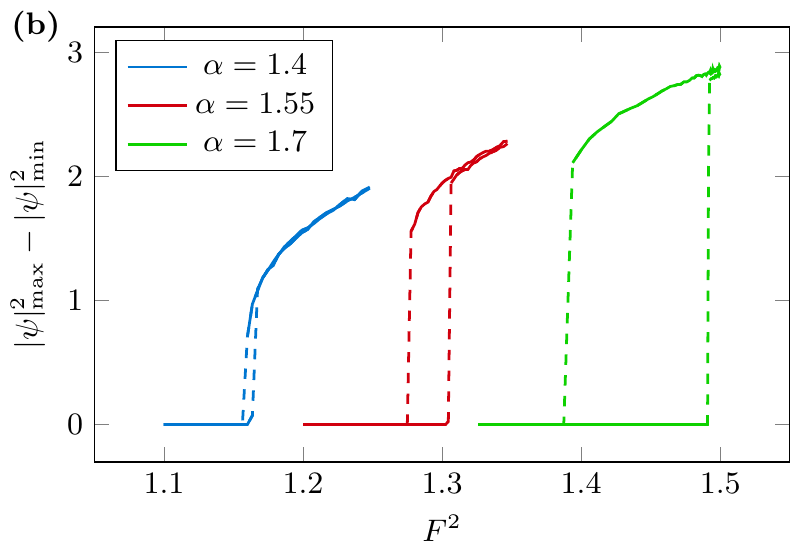}
\end{center}
\caption[num_sub_super]
{\label{num_sub_super} (Color online) Numerical simulations showing the super-critical and sub-critical nature of the Turing patterns as the detuning parameter
$\alpha$ is varied across the critical value $\alpha_{\rm cr} = 41/30$. The other parameters are $\beta= -0.04$ and $\rho = 1.2$.
(a) Growth of the pattern for the super-critical case, where $\alpha < \alpha_{\rm cr}$. It can be seen that after a critical value given by $F_{\rm th}^2 = 1 + (1- \alpha)^2$, the amplitude of the pattern grows smoothly.
(b) Growth of the pattern in the sub-critical case, with $\alpha > \alpha_{\rm cr}$. Here, the hysteresis area can clearly be identified when $\alpha$ is first smoothly swept upwards, and then downwards.  Note that the hysteresis area increases with the detuning $\alpha -\alpha_{\rm cr}$.
}
\end{figure}
%%%%%%%%%%%%%%%%%%%%%%%%%%%%%%%%%%%%%%%%%%%%%%%%%%%%%%%%%%%%%%%%%%%%%%%%%%%%%%%%%%%%%%%%%%%%%%%%%%%%%%%%%%%%%%%%%%%%%%%%%%%%%%%%%%%%%%%%%%%%%%%%%%%%%%%%%%%
%%%%%%%%%%%%%%%%%%%%%%%%%%%%%%%%%%%%%%%%%%%%%%%%%%%%%%%%%%%%%%%%%%%%%%%%%%%%%%%%%%%%%%%%%%%%%%%%%%%%%%%%%%%%%%%%%%%%%%%%%%%%%%%%%%%%%%%%%%%%%%%%%%%%%%%%%%%
%%%%%%%%%%%%%%%%%%%%%%%%%%%%%%%%%%%%%%%%%%%%%%%%%%%%%%%%%%%%%%%%%%%%%%%%%%%%%%%%%%%%%%%%%%%%%%%%%%%%%%%%%%%%%%%%%%%%%%%%%%%%%%%%%%%%%%%%%%%%%%%%%%%%%%%%%%%

Note that in the pictograms of Type~3 and~6, some eigenvalues are located on the imaginary axis, leading to the so-called $(i \omega_1)(i \omega_2)$ and $(i \omega)$ bifurcations respectively. 
However, because of the $\theta \rightarrow -\theta$ reversibility of the system, these eigenvalues stay on the imaginary axis and impede the transversality condition to be fulfilled. These bifurcations therefore dynamically irrelevant in our system and have not been highlighted in Table~\ref{NomenclatureBifurcations}~\cite{Part_I}.

These four bifurcations are exactly those that were obtained in the normal GVD case.
However, the eigenvalue structure in the anomalous GVD case is totally different: all the eigenvalues are rotated by $90^\circ$, because the eigenvalues of the normal and anomalous GVD regimes only differ by a multiplicative factor $i= e^{i\frac{\pi}{4}}$. 

This \textit{essential} difference is the one that explains the intrinsically different dynamics that can be witnessed in both dispersion regimes: very limited in the normal dispersion case, and very rich when the dispersion is anomalous.
For example, it was shown in Part~I~\cite{Part_I} that no Kerr comb generation was possible for $\alpha < \sqrt{3}$ because the line  ${\mathsf{A}}_1$ had eigenvalues of Type~8, which does not correspond to a  bifurcation.
However, in the anomalous GVD regime, this same line corresponds to eigenvalues of Type~7, that is, to a $(i \omega)^2$ bifurcation. This bifurcation at $\rho  = 1 = |\psi_{\rm th}|^2$ has been studied in much detail in ref.~\cite{YanneNanPRA} using a modal expansion approach, and it was shown that it corresponded to the generation of the so called primary comb.
Further analysis shows that this bifurcation corresponds in fact to modulational instability and leads to azimuthal Turing patterns~\cite{IEEE_PJ}.

From a more general perspective, it appears that at the opposite of what was observed in the normal GVD regime, Kerr combs can emerge in the anomalous GVD regime for any value of the cavity detuning $\alpha$. 
The variety of solutions that can be obtained depending on the parameters and on the initial conditions is analyzed in detail in the following sections.

%%%%%%%%%%%%%%%%%%%%%%%%%%%%%%%%%%%%%%%%%%%%%%%%%%%%%%%%%%%%%%%%%%%%%%%%%%%%%%%%%%%%%%%%%%%%%%%%%%%%%%%%%%%%%%%%%%%%%%%%%%%%%%%%%%%%%%%%%%%%%%%%%%%%%%%%%%%
%%%%%%%%%%%%%%%%%%%%%%%%%%%%%%%%%%%%%%%%%%%%%%%%%%%%%%%%%%%%%%%%%%%%%%%%%%%%%%%%%%%%%%%%%%%%%%%%%%%%%%%%%%%%%%%%%%%%%%%%%%%%%%%%%%%%%%%%%%%%%%%%%%%%%%%%%%%
%%%%%%%%%%%%%%%%%%%%%%%%%%%%%%%%%%%%%%%%%%%%%%%%%%%%%%%%%%%%%%%%%%%%%%%%%%%%%%%%%%%%%%%%%%%%%%%%%%%%%%%%%%%%%%%%%%%%%%%%%%%%%%%%%%%%%%%%%%%%%%%%%%%%%%%%%%%
\begin{figure}
\begin{center}
\includegraphics[width=7cm]{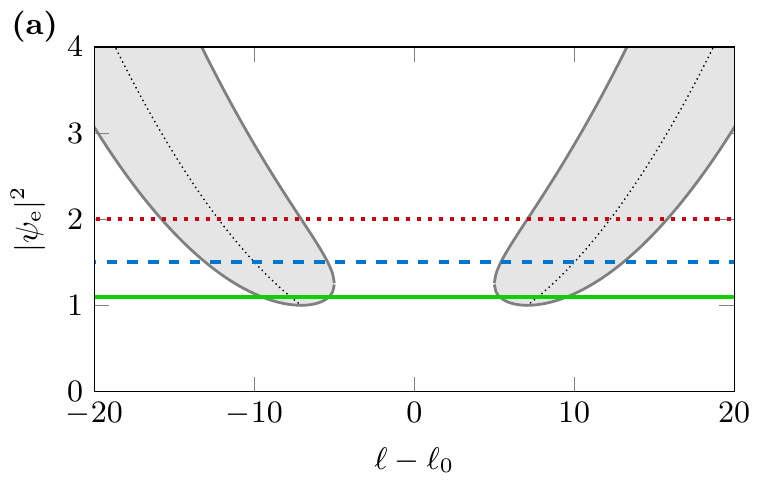}
\includegraphics[width=7cm]{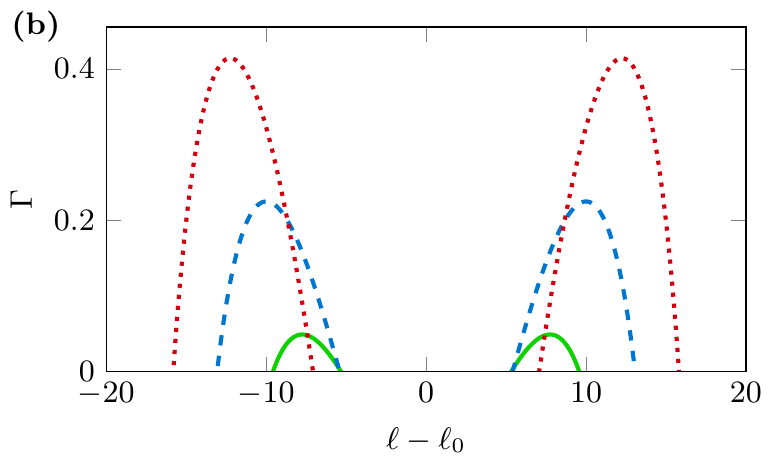}
\end{center}
\caption[MI_Gain]
{\label{MI_Gain} (Color online) 
Determination of the parametric (or MI) gain in the system, with the parameters $\alpha = 1$ and $\beta= - 0.04$.
(a) Positive gain is experienced in the system when the figurative point is in the shaded area of the plane $l$--$\rho$, and negative gain is experienced in the white area. The densely dotted line within the shaded area indicate the location of the maximum gain.
For a fixed value of $\rho$ (that is, of the pump power $F^2$), the corresponding horizontal line intersects the gain area when $\rho >1$, and thereby delimit the modes $ \pm l$ which can grow through MI.
(b) MI gain corresponding to the three pump levels of (a). Note that as the pump power is increased, the maximum gain mode is shifted away from the pump, and the MI gain bandwidth is both shifted outwards and increased as well.   
 }
\end{figure}
%%%%%%%%%%%%%%%%%%%%%%%%%%%%%%%%%%%%%%%%%%%%%%%%%%%%%%%%%%%%%%%%%%%%%%%%%%%%%%%%%%%%%%%%%%%%%%%%%%%%%%%%%%%%%%%%%%%%%%%%%%%%%%%%%%%%%%%%%%%%%%%%%%%%%%%%%%%
%%%%%%%%%%%%%%%%%%%%%%%%%%%%%%%%%%%%%%%%%%%%%%%%%%%%%%%%%%%%%%%%%%%%%%%%%%%%%%%%%%%%%%%%%%%%%%%%%%%%%%%%%%%%%%%%%%%%%%%%%%%%%%%%%%%%%%%%%%%%%%%%%%%%%%%%%%%
%%%%%%%%%%%%%%%%%%%%%%%%%%%%%%%%%%%%%%%%%%%%%%%%%%%%%%%%%%%%%%%%%%%%%%%%%%%%%%%%%%%%%%%%%%%%%%%%%%%%%%%%%%%%%%%%%%%%%%%%%%%%%%%%%%%%%%%%%%%%%%%%%%%%%%%%%%%

\section{Turing patterns}
\label{turingpatterns}

Pattern formation in systems dynamically described by partial differential equations 
was investigated for the first time by Alan Turing 
in his seminal work of morphogenesis~\cite{Turing}.

In our system, the so-called \textit{Turing patterns}  
originate from the $(i \omega)^2$ bifurcation arising at $\rho=1$ for $\alpha <2$
(lines ${\mathsf{A}}_1$ and ${\mathsf{A}}_2$).

In this section, we will mainly focus for the sake of simplification on the case $\alpha < \sqrt{3}$ where this bifurcation is the only one that can occur in the system (portion of line ${\mathsf{A}}_1$).
The emergence of Turing patterns has been studied in ref.~\cite{YanneNanPRA} with a modal expansion approach (where they induced the so-called \textit{primary combs}), and compared to experimental measurements in ref.~\cite{IEEE_PJ}.
In the general case of pattern formation in LLE equations, an abundant literature is indeed available and include for example refs.~\cite{Lugiato_JQE_special,review_springer} (and references therein), and refs.~\cite{Scroggie,IFISC1,2D_solitons,IFISC2} for example focus on the special case of spatial solitons where the laplacian term stands for diffraction instead of dispersion.

In our case, it can be shown that Turing patterns emerge following two different scenarios, either following a supercritical bifurcation (soft excitation), or a sub-critical bifurcation (hard excitation), which are analyzed hereafter.

%%%%%%%%%%%%%%%%%%%%%%%%%%%%%%%%%%%%%%%%%%%%%%%%%%%%%%%%%%%%%%%%%%%%%%%%%%%%%%%%%%%%%%%%%%%%%%%%%%%%%%%%%%%%%%%%%%%%%%%%%%%%%%%%%%%%%%%%%%%%%%%%%%%%%%%%%%%
%%%%%%%%%%%%%%%%%%%%%%%%%%%%%%%%%%%%%%%%%%%%%%%%%%%%%%%%%%%%%%%%%%%%%%%%%%%%%%%%%%%%%%%%%%%%%%%%%%%%%%%%%%%%%%%%%%%%%%%%%%%%%%%%%%%%%%%%%%%%%%%%%%%%%%%%%%%
%%%%%%%%%%%%%%%%%%%%%%%%%%%%%%%%%%%%%%%%%%%%%%%%%%%%%%%%%%%%%%%%%%%%%%%%%%%%%%%%%%%%%%%%%%%%%%%%%%%%%%%%%%%%%%%%%%%%%%%%%%%%%%%%%%%%%%%%%%%%%%%%%%%%%%%%%%%
\begin{figure}
\begin{center}
\includegraphics[width=4.6cm]{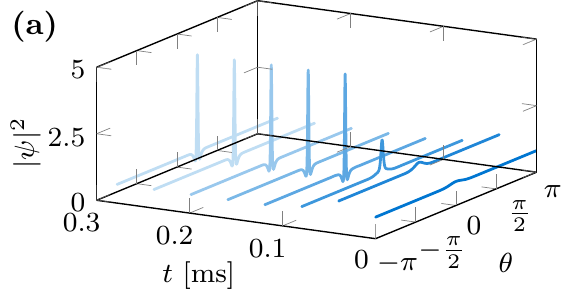}
\includegraphics[width=3.4cm]{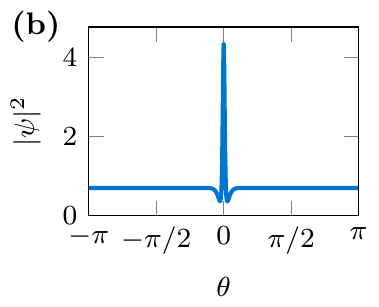}
\includegraphics[width=3.4cm]{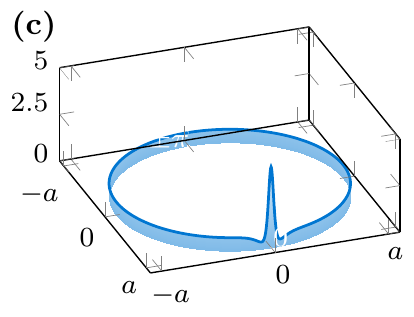}
\includegraphics[width=4.6cm]{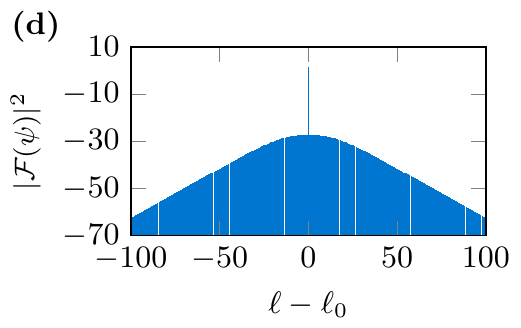}
\end{center}
\caption[Single-soliton]
{\label{Single-soliton} (Color online) Formation of bright solitons.
The parameters are $\alpha =2$, $\beta= -0.004$, and $\rho = 0.7$.
(a) Transient dynamics.
(b) Final pattern in the azimuthal direction.
(c) 3D representation.
(d) Corresponding Kerr comb.
 }
\end{figure}
%%%%%%%%%%%%%%%%%%%%%%%%%%%%%%%%%%%%%%%%%%%%%%%%%%%%%%%%%%%%%%%%%%%%%%%%%%%%%%%%%%%%%%%%%%%%%%%%%%%%%%%%%%%%%%%%%%%%%%%%%%%%%%%%%%%%%%%%%%%%%%%%%%%%%%%%%%%
%%%%%%%%%%%%%%%%%%%%%%%%%%%%%%%%%%%%%%%%%%%%%%%%%%%%%%%%%%%%%%%%%%%%%%%%%%%%%%%%%%%%%%%%%%%%%%%%%%%%%%%%%%%%%%%%%%%%%%%%%%%%%%%%%%%%%%%%%%%%%%%%%%%%%%%%%%%
%%%%%%%%%%%%%%%%%%%%%%%%%%%%%%%%%%%%%%%%%%%%%%%%%%%%%%%%%%%%%%%%%%%%%%%%%%%%%%%%%%%%%%%%%%%%%%%%%%%%%%%%%%%%%%%%%%%%%%%%%%%%%%%%%%%%%%%%%%%%%%%%%%%%%%%%%%%

\subsection{Supercritical and subcritical Turing patterns}
\label{prim_sec_combs}

The super- of subcritical nature if the Turing patterns originating form the LLE was already foreshadowed in the original work of Lugiato and Lefever~\cite{LL}. It has later on been studied extensively by several research groups investigating dissipative structures in nonlinear optical cavities, and two recent noteworthy works on this topic are refs.~\cite{Japanese_Group,Kozyreff}.
The essential difference between a super- and a subcritical pitchfork in our context is explained in Fig.~\ref{super_sub}, and it depends on how the comb emerges around the threshold pump power
\begin{eqnarray}
F_{\rm th}^2 = 1 +(1-\alpha)^2
\label{def_Fth}
\end{eqnarray}
which is obtained from Eq.~(\ref{def_F2}) by setting $\rho = |\psi_{\rm th}|^2 = 1$. 
The numerical simulation of the LLE shows in Fig.~\ref{num_sub_super} that effectively, as the critical value $\alpha_{\rm cr} = 41/30$ is crossed, the growth of the Turing rolls undergoes a structural change which is mathematically explained by the paradigm of super- and sub-critical pitchfork bifurcations.

%%%%%%%%%%%%%%%%%%%%%%%%%%%%%%%%%%%%%%%%%%%%%%%%%%%%%%%%%%%%%%%%%%%%%%%%%%%%%%%%%%%%%%%%%%%%%%%%%%%%%%%%%%%%%%%%%%%%%%%%%%%%%%%%%%%%%%%%%%%%%%%%%%%%%%%%%%%
%%%%%%%%%%%%%%%%%%%%%%%%%%%%%%%%%%%%%%%%%%%%%%%%%%%%%%%%%%%%%%%%%%%%%%%%%%%%%%%%%%%%%%%%%%%%%%%%%%%%%%%%%%%%%%%%%%%%%%%%%%%%%%%%%%%%%%%%%%%%%%%%%%%%%%%%%%%
%%%%%%%%%%%%%%%%%%%%%%%%%%%%%%%%%%%%%%%%%%%%%%%%%%%%%%%%%%%%%%%%%%%%%%%%%%%%%%%%%%%%%%%%%%%%%%%%%%%%%%%%%%%%%%%%%%%%%%%%%%%%%%%%%%%%%%%%%%%%%%%%%%%%%%%%%%%
\begin{figure}
\begin{center}
\includegraphics[width=4.6cm]{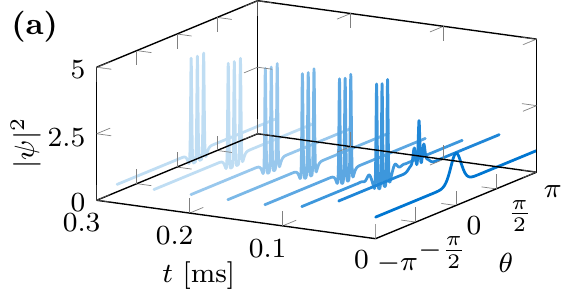}
\includegraphics[width=3.4cm]{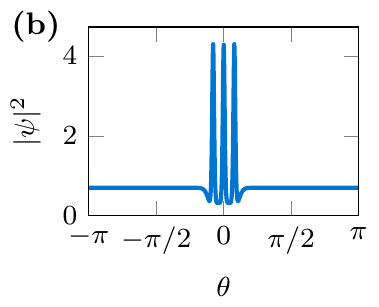}
\includegraphics[width=4cm]{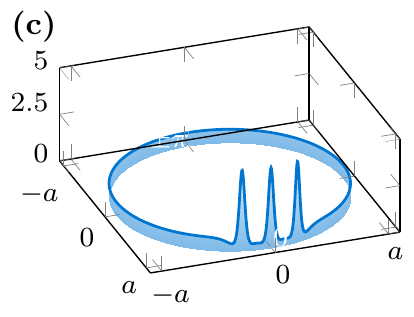}
\includegraphics[width=4cm]{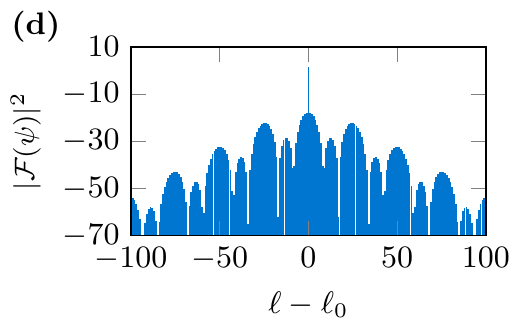}
\includegraphics[width=4cm]{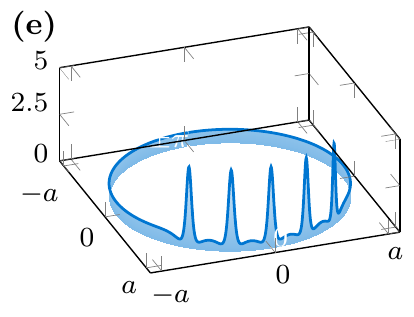}
\includegraphics[width=4cm]{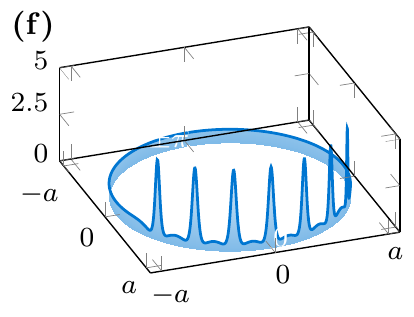}
\end{center}
\caption[soliton_molecule]
{\label{soliton_molecule} (Color online) Formation of a soliton molecule.
The parameters are exactly those of Fig.~\ref{Single-soliton}, and only the initial condition has changed (more powerful pulse).
It can be inferred that a soliton is in fact a pulse isolated from a sub-critical Turing pattern, and that a soliton molecule corresponds to several of such pulses.
(a) Transient dynamics.
(b) Final pattern in the azimuthal direction.
(c) 3D representation.
(d) Corresponding Kerr comb.
(e) Soliton molecule formed with 5~solitons obtained with a different initial condition.
(f) Soliton molecule formed with 7~solitons obtained with a different initial condition.
 }
\end{figure}
%%%%%%%%%%%%%%%%%%%%%%%%%%%%%%%%%%%%%%%%%%%%%%%%%%%%%%%%%%%%%%%%%%%%%%%%%%%%%%%%%%%%%%%%%%%%%%%%%%%%%%%%%%%%%%%%%%%%%%%%%%%%%%%%%%%%%%%%%%%%%%%%%%%%%%%%%%%
%%%%%%%%%%%%%%%%%%%%%%%%%%%%%%%%%%%%%%%%%%%%%%%%%%%%%%%%%%%%%%%%%%%%%%%%%%%%%%%%%%%%%%%%%%%%%%%%%%%%%%%%%%%%%%%%%%%%%%%%%%%%%%%%%%%%%%%%%%%%%%%%%%%%%%%%%%%
%%%%%%%%%%%%%%%%%%%%%%%%%%%%%%%%%%%%%%%%%%%%%%%%%%%%%%%%%%%%%%%%%%%%%%%%%%%%%%%%%%%%%%%%%%%%%%%%%%%%%%%%%%%%%%%%%%%%%%%%%%%%%%%%%%%%%%%%%%%%%%%%%%%%%%%%%%%

On the one hand, a supercritical bifurcation to Turing patterns occurs when $\alpha < 41/30$. 
In this case, the unique equilibrium $\psi_{\rm e}$ is stable when below the threshold pump power $F < F_{\rm th}$ and unstable when $F > F_{\rm th}$, leading to the emergence of the Turing pattern (primary Kerr comb in the spectral domain). 
Above the pump threshold $F_{\rm th}$, the sidemodes of the Kerr comb can continuously grow from infinitesimally small to significantly large as the pump is increased. However, the equilibrium and the Turing pattern are never simultaneously stable. 
The formation of these supercritical patterns is displayed in Fig.~\ref{Supercritical_Turing}.
It is interesting to note that the rolls are smooth, and yield a Kerr comb characterized by isolated spectral lines with multiple FSR separation, the multiplicity being equal to the number of rolls.

On the other hand, a sub-critical  bifurcation to Turing patterns arises when $\alpha > 41/30$.
Here, as for the supercritical case, the equilibrium $\psi_{\rm e}$ is stable for $F < F_{\rm th}$ and unstable above.
However, Kerr comb is possible in a small range \emph{below} the threshold pump power $F_{\rm th}^2$.
Hence, in the small interval below $F_{\rm th}^2$ where the comb can be excited, a stable Turing pattern does coexist with the equilibrium solution. This situation creates a bistability and also induces hysteresis, as the dynamical state of the system will not be the same if the pump is adiabatically increased comparatively to when it is decreased.
The consequence of this bistability is that the transition from the equilibrium to the Turing pattern is abrupt, and the sidemodes of the Kerr comb can not be infinitesimally small as it was the case for the supercritical case. 
Figure~\ref{Subcritical_Turing} shows the formation of these sub-critical patterns. At the opposite of the super-critical rolls, small pedestals can be observed in this case, and the rolls also appear to be sharper.

In the context of Kerr comb generation, the super- and sub-critical bifurcation have sometimes been referred to as \textit{soft} and \textit{hard} excitation modes, respectively~\cite{Matsko_soft_hard}. It is also noteworthy that the Turing patterns beyond $\alpha =\sqrt{3}$ are still sub-critical: however, the eigenvalue structure becomes more complex because this area in the parameter space $\alpha$--$F^2$ can encompass multiple-equilibria and the four types of bifurcations listed in Sec.~\ref{bifurcationmap}. This is the area where bright cavity solitons and related structures can emerge. We will study these complex sub-critical patterns the next sections.

%%%%%%%%%%%%%%%%%%%%%%%%%%%%%%%%%%%%%%%%%%%%%%%%%%%%%%%%%%%%%%%%%%%%%%%%%%%%%%%%%%%%%%%%%%%%%%%%%%%%%%%%%%%%%%%%%%%%%%%%%%%%%%%%%%%%%%%%%%%%%%%%%%%%%%%%%%%
%%%%%%%%%%%%%%%%%%%%%%%%%%%%%%%%%%%%%%%%%%%%%%%%%%%%%%%%%%%%%%%%%%%%%%%%%%%%%%%%%%%%%%%%%%%%%%%%%%%%%%%%%%%%%%%%%%%%%%%%%%%%%%%%%%%%%%%%%%%%%%%%%%%%%%%%%%%
%%%%%%%%%%%%%%%%%%%%%%%%%%%%%%%%%%%%%%%%%%%%%%%%%%%%%%%%%%%%%%%%%%%%%%%%%%%%%%%%%%%%%%%%%%%%%%%%%%%%%%%%%%%%%%%%%%%%%%%%%%%%%%%%%%%%%%%%%%%%%%%%%%%%%%%%%%%
\begin{figure}
\begin{center}
\includegraphics[width=7cm]{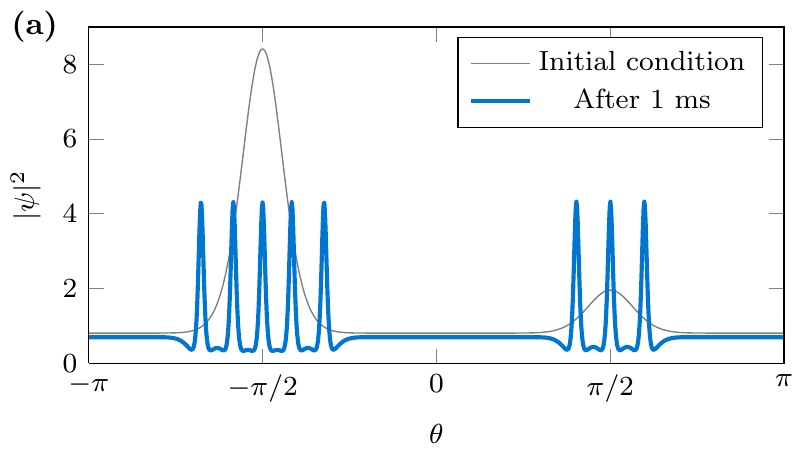}
\includegraphics[width=7cm]{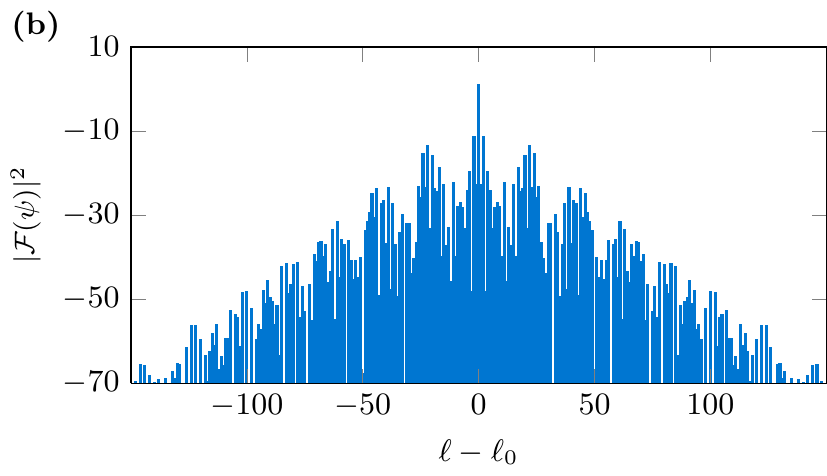}
\end{center}
\caption[coexistence_molecules]
{\label{coexistence_molecules} (Color online) 
Coexistence of two soliton molecules in the resonator.
The parameters are those of Fig.~\ref{Single-soliton}.
(a) Initial conditions and final state.
(b) Corresponding Kerr comb.
 }
\end{figure}
%%%%%%%%%%%%%%%%%%%%%%%%%%%%%%%%%%%%%%%%%%%%%%%%%%%%%%%%%%%%%%%%%%%%%%%%%%%%%%%%%%%%%%%%%%%%%%%%%%%%%%%%%%%%%%%%%%%%%%%%%%%%%%%%%%%%%%%%%%%%%%%%%%%%%%%%%%%
%%%%%%%%%%%%%%%%%%%%%%%%%%%%%%%%%%%%%%%%%%%%%%%%%%%%%%%%%%%%%%%%%%%%%%%%%%%%%%%%%%%%%%%%%%%%%%%%%%%%%%%%%%%%%%%%%%%%%%%%%%%%%%%%%%%%%%%%%%%%%%%%%%%%%%%%%%%
%%%%%%%%%%%%%%%%%%%%%%%%%%%%%%%%%%%%%%%%%%%%%%%%%%%%%%%%%%%%%%%%%%%%%%%%%%%%%%%%%%%%%%%%%%%%%%%%%%%%%%%%%%%%%%%%%%%%%%%%%%%%%%%%%%%%%%%%%%%%%%%%%%%%%%%%%%%

\subsection{Number of rolls in the Turing patterns}
\label{number_rolls}

The number of rolls in the Turing pattern arising from the $(i \omega)^2$ bifurcation at $\rho=1$ necessarily requires to account for the boundary conditions. The reason is that in this case, the patterns fill the whole $\theta$-domain and the number of rolls along the azimuthal direction has to be an integer. Hence, the acknowledging for the modal structure of the patterns is there particularly relevant to understand this phenomenology.

%%%%%%%%%%%%%%%%%%%%%%%%%%%%%%%%%%%%%%%%%%%%%%%%%%%%%%%%%%%%%%%%%%%%%%%%%%%%%%%%%%%%%%%%%%%%%%%%%%%%%%%%%%%%%%%%%%%%%%%%%%%%%%%%%%%%%%%%%%%%%%%%%%%%%%%%%%%
%%%%%%%%%%%%%%%%%%%%%%%%%%%%%%%%%%%%%%%%%%%%%%%%%%%%%%%%%%%%%%%%%%%%%%%%%%%%%%%%%%%%%%%%%%%%%%%%%%%%%%%%%%%%%%%%%%%%%%%%%%%%%%%%%%%%%%%%%%%%%%%%%%%%%%%%%%%
%%%%%%%%%%%%%%%%%%%%%%%%%%%%%%%%%%%%%%%%%%%%%%%%%%%%%%%%%%%%%%%%%%%%%%%%%%%%%%%%%%%%%%%%%%%%%%%%%%%%%%%%%%%%%%%%%%%%%%%%%%%%%%%%%%%%%%%%%%%%%%%%%%%%%%%%%%%
\begin{figure}
\begin{center}
\includegraphics[width=4.6cm]{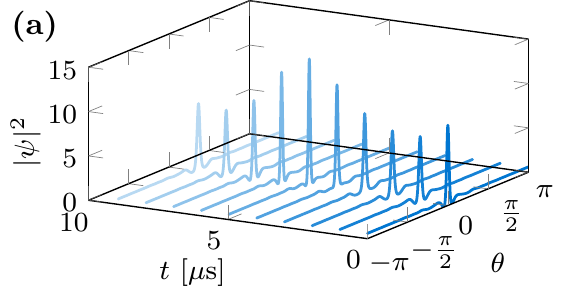}
\includegraphics[width=3.4cm]{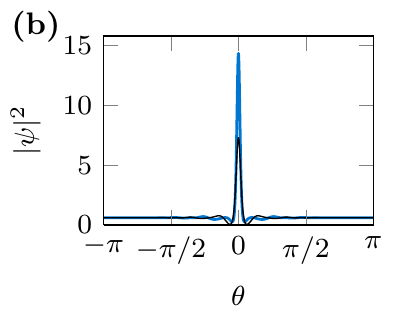}
\includegraphics[width=4cm]{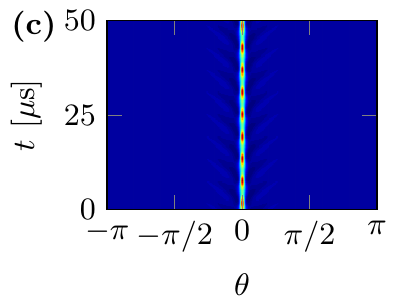}
\includegraphics[width=4cm]{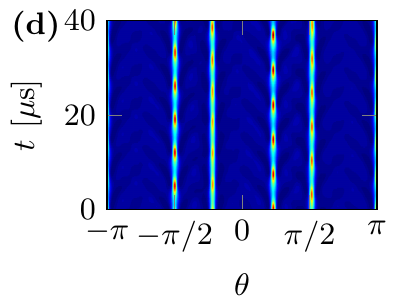}
\end{center}
\caption[breathers]
{\label{breathers} (Color online) Soliton breathers.
The parameters are $\alpha =4$, $\beta= -0.04$, and $\rho = 0.7$ ($F^2= 7.54$).
(a) Time-domain dynamics.
(b) Maximal and minimal pulse-shapes.
(c) Color-coded visualization of the time-domain dynamics of a soliton breather.
(d) Color-coded visualization of the time-domain dynamics of a complex structure corresponding to a higher-order soliton breather.
}
\end{figure}
%%%%%%%%%%%%%%%%%%%%%%%%%%%%%%%%%%%%%%%%%%%%%%%%%%%%%%%%%%%%%%%%%%%%%%%%%%%%%%%%%%%%%%%%%%%%%%%%%%%%%%%%%%%%%%%%%%%%%%%%%%%%%%%%%%%%%%%%%%%%%%%%%%%%%%%%%%%
%%%%%%%%%%%%%%%%%%%%%%%%%%%%%%%%%%%%%%%%%%%%%%%%%%%%%%%%%%%%%%%%%%%%%%%%%%%%%%%%%%%%%%%%%%%%%%%%%%%%%%%%%%%%%%%%%%%%%%%%%%%%%%%%%%%%%%%%%%%%%%%%%%%%%%%%%%%
%%%%%%%%%%%%%%%%%%%%%%%%%%%%%%%%%%%%%%%%%%%%%%%%%%%%%%%%%%%%%%%%%%%%%%%%%%%%%%%%%%%%%%%%%%%%%%%%%%%%%%%%%%%%%%%%%%%%%%%%%%%%%%%%%%%%%%%%%%%%%%%%%%%%%%%%%%%

According to the LLE, a perturbation $\delta \psi (\theta, \tau)$ of the equilibrium (flat solution) $\psi_{\rm e}$ obeys the linearized equation 
\begin{eqnarray}
\frac{\partial }{\partial \tau} [\delta \psi] &=&  - (1 +i\alpha)  \delta \psi 
                                                  + 2 i |\psi_{\rm e}|^2 \delta \psi 
                                                + i \psi_{\rm e}^2 \delta \psi^* \nonumber \\
                                              && -i \frac{\beta}{2} \frac{\partial^2}{\partial \theta^2}[\delta \psi] \, .
\label{perturb_LLE}
\end{eqnarray}
Following Eq.~(\ref{psi_expansion}), we can expand this perturbation according to the ansatz
\begin{eqnarray}
\delta \psi  (\theta, \tau) = \sum_l a_l(\tau) \, e^{i l \theta}
\label{ansatz_perturbation}
\end{eqnarray}
where $l \equiv \ell -\ell_0$ corresponds to the eigennumber of the WGMs with respect to the pumped mode $\ell_0$.
The interacting eigenmodes can therefore be synthetically labelled as $\pm 1$, $\pm 2$, $\cdots$, the mode $l=0$ being the central mode. 
After inserting the ansatz of Eq.~(\ref{ansatz_perturbation}) into Eq.~(\ref{perturb_LLE}),  we obtain an equation which can be used to perform an hermitian projection in order to track the individual dynamics of the modal perturbations $a_l$.
A projection onto a given mode $l'$  consists in multiplying the equation by  $e^{i l' \theta}$, and integrating the product from $-\pi$ to $\pi$ with respect to $\theta$.
The result of this projection yields two equations for the modal perturbations, which appear to be pairwise coupled according to 
 \begin{eqnarray}
\left[ \begin{array}{l}
		\dot{a}_{l} \\
		\dot{a}_{-l}^*
	  \end{array}
\right]
=
\left[ \begin{array}{ll}
      {\cal M}  & {\cal N} \\
      {\cal N}^* & {\cal M}^*
	  \end{array}
\right] 
\left[ \begin{array}{l}
		a_{l} \\
		a_{-l}^*
	  \end{array}
\right]
\, ,
\label{eq_matrix_perturb_a}
\end{eqnarray}	 
where the overdot stands for the derivative with respect to the dimensionless time $\tau$, while 
\begin{eqnarray}
{\cal M} &=&  - (1+i \alpha) + 2 i |\psi_{\rm e}|^2 + i \frac{\beta}{2} l^2  \nonumber \\
{\cal N} &=&   i \psi_{\rm e}^2  \, . 
\label{eq_MN}
\end{eqnarray}
The eigenvalues of the matrix in Eq.~(\ref{eq_matrix_perturb_a}) define if a small signal perturbation (noise) in the modes  ${\pm l}$ increases or decreases with time. In particular, the real part of the leading eigenvalue (the one with the largest real part) can be viewed as the a \emph{gain} parameter, which can be explicitly written as 
\begin{eqnarray}
\Gamma (l) = \Re \Bigg\{ -1 + \sqrt{\rho - \frac{1}{4} \left[ \alpha - 2 \rho - \frac{1}{2} \beta l^2 \right]^2} \Bigg\} \, ,
\label{eq_gain}
\end{eqnarray}
where $\rho = |\psi_{\rm e}|^2$.

%%%%%%%%%%%%%%%%%%%%%%%%%%%%%%%%%%%%%%%%%%%%%%%%%%%%%%%%%%%%%%%%%%%%%%%%%%%%%%%%%%%%%%%%%%%%%%%%%%%%%%%%%%%%%%%%%%%%%%%%%%%%%%%%%%%%%%%%%%%%%%%%%%%%%%%%%%%
%%%%%%%%%%%%%%%%%%%%%%%%%%%%%%%%%%%%%%%%%%%%%%%%%%%%%%%%%%%%%%%%%%%%%%%%%%%%%%%%%%%%%%%%%%%%%%%%%%%%%%%%%%%%%%%%%%%%%%%%%%%%%%%%%%%%%%%%%%%%%%%%%%%%%%%%%%%
%%%%%%%%%%%%%%%%%%%%%%%%%%%%%%%%%%%%%%%%%%%%%%%%%%%%%%%%%%%%%%%%%%%%%%%%%%%%%%%%%%%%%%%%%%%%%%%%%%%%%%%%%%%%%%%%%%%%%%%%%%%%%%%%%%%%%%%%%%%%%%%%%%%%%%%%%%%
\begin{figure}
\begin{center}
\includegraphics[width=3.4cm]{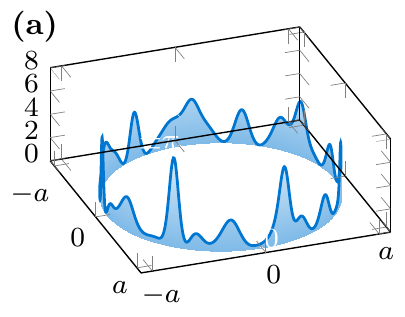}
\includegraphics[width=4.6cm]{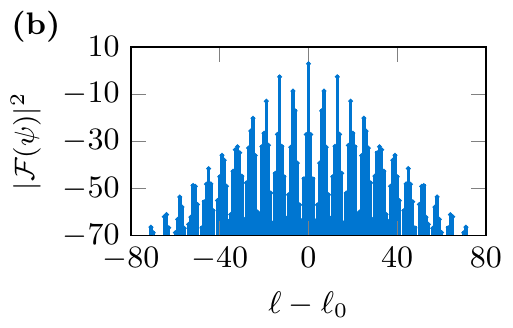}
\includegraphics[width=3.4cm]{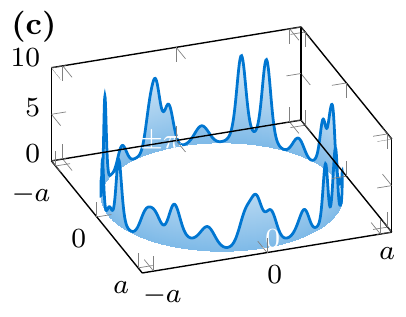}
\includegraphics[width=4.6cm]{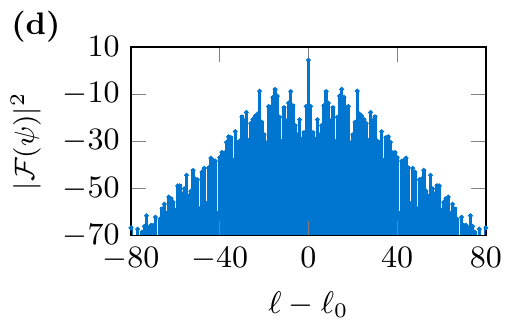}
\includegraphics[width=3.4cm]{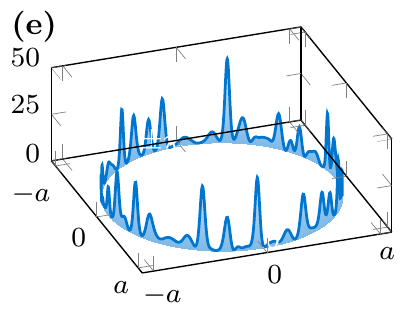}
\includegraphics[width=4.6cm]{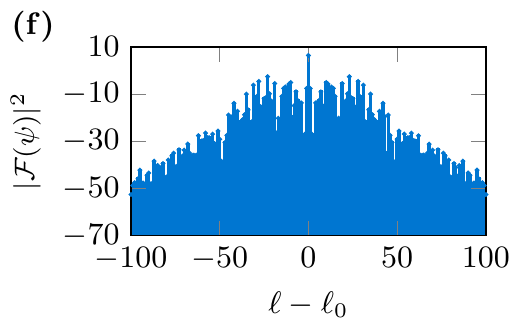}
\end{center}
\caption[chaos]
{\label{chaos} (Color online) Chaos. 
(a) 3D snapshot of a chaotic state for $\alpha =0$, $\beta= -0.04$, and $\rho = 1.9$.
(b) Corresponding Kerr comb.
(c) 3D snapshot of a chaotic state for $\alpha =0$, $\beta= -0.04$, and $\rho = 2.5$.
(d) Corresponding Kerr comb.
(e) 3D snapshot of a chaotic state for $\alpha =2$, $\beta= -0.04$, and $\rho = 3$.
(f) Corresponding Kerr comb.
 }
\end{figure}
%%%%%%%%%%%%%%%%%%%%%%%%%%%%%%%%%%%%%%%%%%%%%%%%%%%%%%%%%%%%%%%%%%%%%%%%%%%%%%%%%%%%%%%%%%%%%%%%%%%%%%%%%%%%%%%%%%%%%%%%%%%%%%%%%%%%%%%%%%%%%%%%%%%%%%%%%%%
%%%%%%%%%%%%%%%%%%%%%%%%%%%%%%%%%%%%%%%%%%%%%%%%%%%%%%%%%%%%%%%%%%%%%%%%%%%%%%%%%%%%%%%%%%%%%%%%%%%%%%%%%%%%%%%%%%%%%%%%%%%%%%%%%%%%%%%%%%%%%%%%%%%%%%%%%%%
%%%%%%%%%%%%%%%%%%%%%%%%%%%%%%%%%%%%%%%%%%%%%%%%%%%%%%%%%%%%%%%%%%%%%%%%%%%%%%%%%%%%%%%%%%%%%%%%%%%%%%%%%%%%%%%%%%%%%%%%%%%%%%%%%%%%%%%%%%%%%%%%%%%%%%%%%%%

At threshold, we have already demonstrated that $\rho=1$. On the other hand, a mode $l$ is excited through MI when it experiences positive gain: the threshold gain can therefore be defined by  $\Gamma =0$.  
Hence, we deduce from the two preceding relations that  at threshold, the two modes $\pm l_{\rm th} $ with 
\begin{eqnarray}
l_{\rm th}  =  \sqrt{\frac{2}{\beta} (\alpha - 2)} \, ,
\label{def_l_th}
\end{eqnarray}
are excited through MI. This number also corresponds to the number of rolls that will be observed in the temporal domain.
This analysis corresponds to the one that has been performed in ref.~\cite{YanneNanPRA} to explain the emergence of the so-called primary comb.

%%%%%%%%%%%%%%%%%%%%%%%%%%%%%%%%%%%%%%%%%%%%%%%%%%%%%%%%%%%%%%%%%%%%%%%%%%%%%%%%%%%%%%%%%%%%%%%%%%%%%%%%%%%%%%%%%%%%%%%%%%%%%%%%%%%%%%%%%%%%%%%%%%%%%%%%%%%
%%%%%%%%%%%%%%%%%%%%%%%%%%%%%%%%%%%%%%%%%%%%%%%%%%%%%%%%%%%%%%%%%%%%%%%%%%%%%%%%%%%%%%%%%%%%%%%%%%%%%%%%%%%%%%%%%%%%%%%%%%%%%%%%%%%%%%%%%%%%%%%%%%%%%%%%%%%
%%%%%%%%%%%%%%%%%%%%%%%%%%%%%%%%%%%%%%%%%%%%%%%%%%%%%%%%%%%%%%%%%%%%%%%%%%%%%%%%%%%%%%%%%%%%%%%%%%%%%%%%%%%%%%%%%%%%%%%%%%%%%%%%%%%%%%%%%%%%%%%%%%%%%%%%%%%
\begin{figure*}
\begin{center}
\includegraphics[width=16cm]{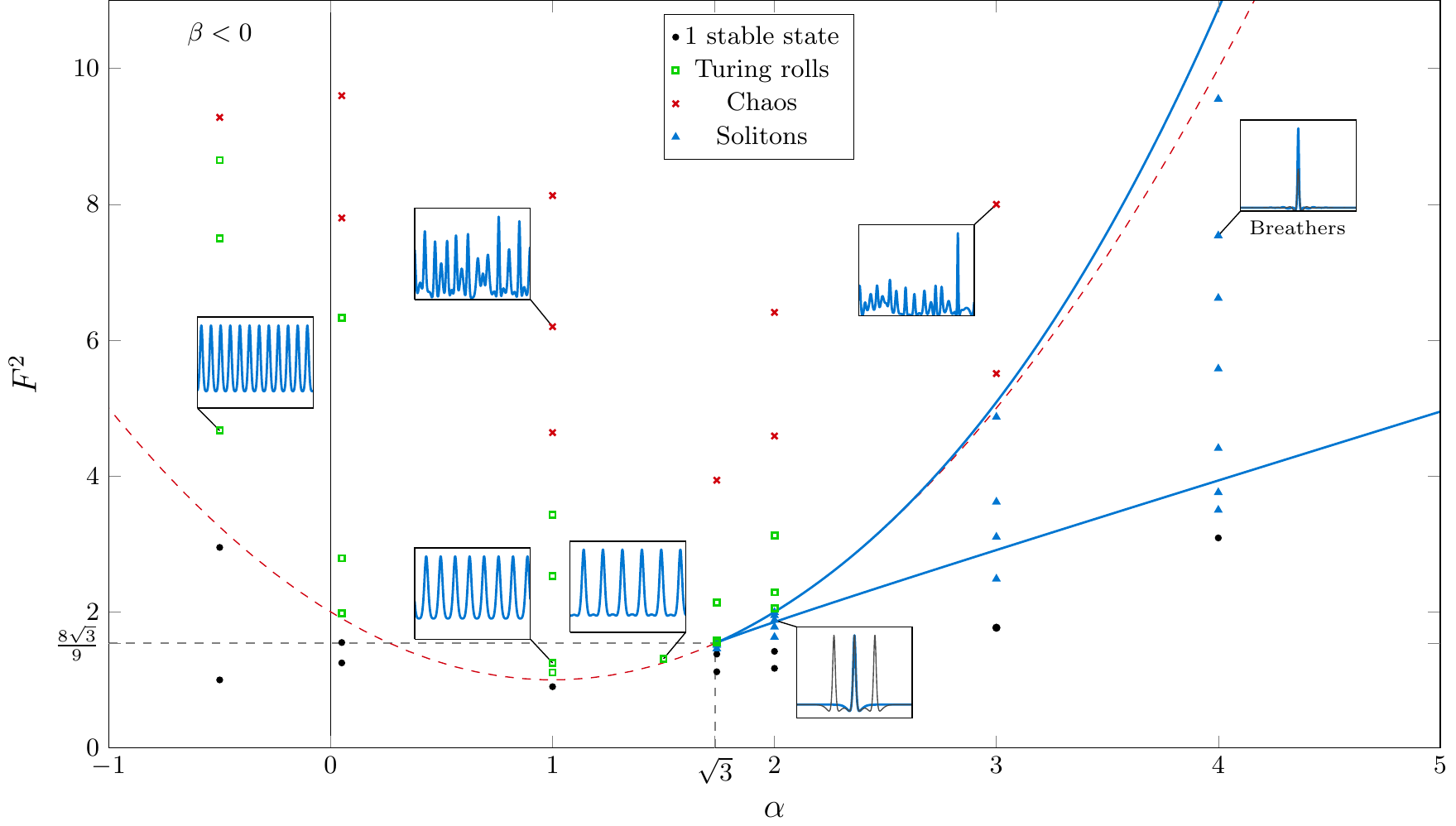}
\end{center}
\caption[Figurebifatscale]
{\label{Figurebifatscale} (Color online) Bifurcation diagram at scale, showing the parameters leading to various stationary solutions. Beside equilibria (flat solutions), the possible solutions are Turing patterns (super- and sub-critical), solitons (and soliton molecules), breathers, and chaos. The connection between solitons and sub-critical Turing patterns appears clearly. Note that solitons and soliton complexes are localized in the same area of this parameter space since only the initial conditions define if the final steady-state will be of one kind or of the other.}
\end{figure*}
%%%%%%%%%%%%%%%%%%%%%%%%%%%%%%%%%%%%%%%%%%%%%%%%%%%%%%%%%%%%%%%%%%%%%%%%%%%%%%%%%%%%%%%%%%%%%%%%%%%%%%%%%%%%%%%%%%%%%%%%%%%%%%%%%%%%%%%%%%%%%%%%%%%%%%%%%%%
%%%%%%%%%%%%%%%%%%%%%%%%%%%%%%%%%%%%%%%%%%%%%%%%%%%%%%%%%%%%%%%%%%%%%%%%%%%%%%%%%%%%%%%%%%%%%%%%%%%%%%%%%%%%%%%%%%%%%%%%%%%%%%%%%%%%%%%%%%%%%%%%%%%%%%%%%%%
%%%%%%%%%%%%%%%%%%%%%%%%%%%%%%%%%%%%%%%%%%%%%%%%%%%%%%%%%%%%%%%%%%%%%%%%%%%%%%%%%%%%%%%%%%%%%%%%%%%%%%%%%%%%%%%%%%%%%%%%%%%%%%%%%%%%%%%%%%%%%%%%%%%%%%%%%%%

From a more general perspective, the modes $l$ that can be directly excited by the pump are such that $\Gamma (l) >0$.
The modes $\pm l_{\rm mgm}$  for which the gain is maximal are referred to as the maximum gain modes (MGM), and they are found through the condition $\partial \Gamma / \partial l =0$, which yields
\begin{eqnarray}
l_{\rm mgm}  =   \sqrt{\frac{2}{\beta} (\alpha - 2 \rho)} \, ,
\label{def_l_mgm}
\end{eqnarray}
Figure~\ref{MI_Gain} graphically displays how the MI gain leads to the emergence of a Turing patterns.
When the system is pumped above threshold ($\rho>1$), two symmetric spectral bands are created around the pump.
The modes that are the most likely to arise from noise are those who have the largest gain, namely $\pm l_{\rm mgm}$.
Near threshold, only two sidemodes (around $ \pm l_{\rm mgm} \simeq \pm l_{\rm th}$) are generated and in the temporal domain: the flat background becomes unstable and leads to the emergence of the rolls, which correspond here to a sinusoidal modulation of the flat solution (from this phenomenology was coined the term modulational instability). 
However, as the pump is increased, $|l_{\rm mgm}|$ increases as well as it can be seen in Fig.~\ref{MI_Gain}(b).
But more importantly, higher-order sidemodes (harmonics) are generated at eigennumbers $ \pm k \times l_{\rm mgm}$, where $k$ is an integer number. As a consequence, the modulation in the time domain is not sinusoidal anymore, but gradually morphs into a train of sharply peaked pulses.

\section{Bright Cavity solitons}
\label{brightcavitysolitons}

The existence of bright cavity solitons in nonlinear optical cavities is a well documented topic.
They arise as a balance between nonlinearity and anomalous dispersion (which defines their shape), 
and a balance between gain and dissipation (which defines their amplitude).
 
Figure~\ref{Single-soliton} displays the transient dynamics towards a cavity soliton.
The initial condition here is a very narrow and small pulse, which grows and converges towards a soliton characterized by a narrow pulse-width and small pedestal oscillations.
This soliton is sub-critical as it emerged for a pump power for which the steady state is such that $\rho <1$: hence, it does not emerge for arbitrarily small (noisy) perturbations of the intra-cavity background field.
In fact, it can be inferred that the soliton is a pulse that has been ``carved out'' of a sub-critical Turing pattern.
This explains at the same time the sub-criticality and the pedestal oscillations, which are indeed observable in sub-critical Turing patterns when the detuning $\alpha$ is not too close to the critical value $41/30$.  
However, at the opposite of Turing patterns, the bright soliton is a \textit{localized structure} in the sense that it does not feel the boundaries when they are at a distance that is significantly larger than its pulse-width.
Hence, the soliton of Fig.~\ref{Single-soliton}(c) dynamically  behaves as if its background had an infinite extension.
The spectrum of this soliton as presented in Fig.~\ref{Single-soliton}(d) has a single-FSR spacing and displays hundreds of mode-locked WGMs. 
Such solitons have also been observed experimentally in recent experiments~\cite{Arxiv_Soliton_Kipp}.
It is also known that the spectral extension of this comb becomes larger as the pulses are narrower: this situation is observed when $\beta \rightarrow 0$ as discussed in Sec.~\ref{influencedispersion}.

\section{Bright Soliton molecules}
\label{brightsolitonsmolecules}

When the parameters lead to the formation of super-critical Turing patterns, the final steady-state is invariably the same patterns regardless of the initial conditions (provided that there is no zero-energy mode at $\tau=0$). 
For sub-critical structures, the situations is indeed very different, as the final output critically depends on the initial conditions.

As far as solitons are concerned, the initial conditions can lead to single-peaked pulses as displayed in
Fig.~\ref{Single-soliton}. However, more energetic initial conditions can lead to the formation of multi-peaked solutions that which are here referred to as \textit{soliton molecules}.
These molecules can be considered has a limited number of pulsed carved out of a sub-critical Turing pattern.
Figure~\ref{soliton_molecule} shows how the three-peaked soliton molecule is formed. 
This dissipative structure is subcritical and localized, exactly as the single soliton.
The corresponding Kerr comb is also characterized by a single-FSR spacing, but in this case, at the opposite of what is observed in the single-peaked soliton case, the spectrum displays a slow modulation.

For a pulse-like excitation, the number of solitons in a molecule can be controlled by the energy 
\begin{eqnarray}
E=\int_{-\pi}^{\pi} |\psi (\theta, \tau=0)|^2 \, d\theta \, , 
\label{energy_IC}
\end{eqnarray}
and it can be shown that the number of solitons in the molecule is governed by a snaking bifurcation.
Figures~\ref{soliton_molecule}(e) and~(f) display soliton molecules with $5$ and~$7$ elements, respectively, and larger numbers can be achieved as long as the molecule is blind to the finiteness of the $\theta$--domain.

It is also noteworthy that different soliton molecules can coexist inside the disk.
Such composite structures can be obtained for example using initial conditions as displayed in Fig.~\ref{coexistence_molecules}(a).
The corresponding Kerr combs look noisy, and might even wrongfully be considered as ``chaotic'': however,  in the time domain, the pattern is perfectly periodic and deterministic. Genuinely chaotic spectra will be studied in Sec.~\ref{Chaos}.

Note that if $N$ individual and non-interacting bright solitons are exactly separated by an angle of $2 \pi/N$ in the $\theta$-domain, the resulting Kerr comb will feature a multiple-FSR structure (exactly like the spectra of Turing rolls). 
These kind of structures were presented in ref.~\cite{Part_I} with dark solitons.

\section{Breathing solitons}
\label{Breathingsolitons}
 
An interesting solution that can be obtained in the LLE is the breather soliton~\cite{Matsko_breathers,Leo_breathers}.
It consists in a soliton whose amplitude varies periodically in time.
However, this period of the breathing is very low, and is of the order of the photon lifetime.
In the spectral domain, the comb corresponding to a single breathing soliton looks like the one of a normal (steady) soliton, except that there are modulation side-bands \textit{inside} the modal resonance linewidths. 

As displayed in Fig.~\ref{breathers}(a) and (b), the breather soliton oscillates in time and it approximately keeps the same pulse-width.
Figure.~\ref{breathers}(c) shows the oscillating behavior at a larger timescale, and it can be observed that this breather soliton is a localized structure that is boundary-blind.
On the other hand, Fig.~\ref{breathers}(d) presents a higher order breather soliton with a complex structure consisting of multiple peaks that are not oscillating in synchrony. Actually, these soliton breathers can have a very wide variety of shapes and oscillation behaviors depending on the initial conditions.

%%%%%%%%%%%%%%%%%%%%%%%%%%%%%%%%%%%%%%%%%%%%%%%%%%%%%%%%%%%%%%%%%%%%%%%%%%%%%%%%%%%%%%%%%%%%%%%%%%%%%%%%%%%%%%%%%%%%%%%%%%%%%%%%%%%%%%%%%%%%%%%%%%%%%%%%%%%
%%%%%%%%%%%%%%%%%%%%%%%%%%%%%%%%%%%%%%%%%%%%%%%%%%%%%%%%%%%%%%%%%%%%%%%%%%%%%%%%%%%%%%%%%%%%%%%%%%%%%%%%%%%%%%%%%%%%%%%%%%%%%%%%%%%%%%%%%%%%%%%%%%%%%%%%%%%
%%%%%%%%%%%%%%%%%%%%%%%%%%%%%%%%%%%%%%%%%%%%%%%%%%%%%%%%%%%%%%%%%%%%%%%%%%%%%%%%%%%%%%%%%%%%%%%%%%%%%%%%%%%%%%%%%%%%%%%%%%%%%%%%%%%%%%%%%%%%%%%%%%%%%%%%%%%
\begin{figure}
\begin{center}
\includegraphics[width=7cm]{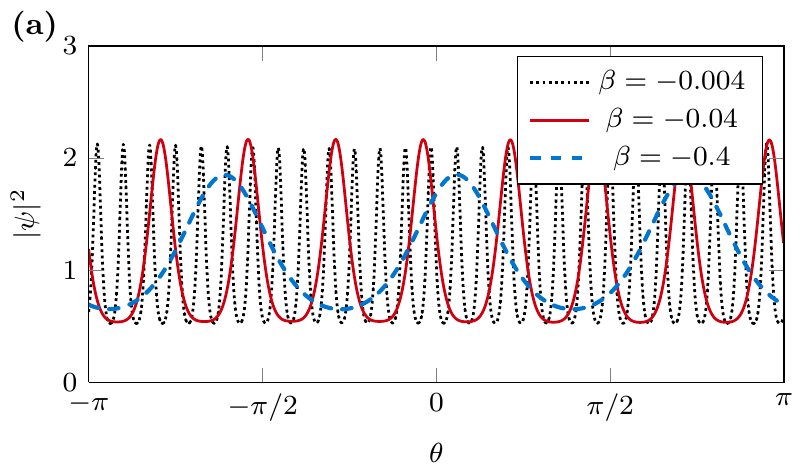}
\includegraphics[width=7cm]{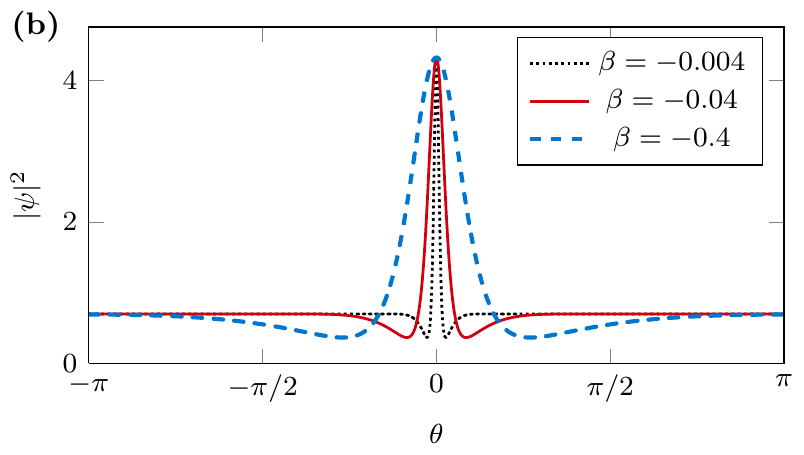}
\end{center}
\caption[influence_beta]
{\label{influence_beta} (Color online) Influence of the magnitude of the dispersion parameter $\beta$.
(a) Case of Turing patterns. The influence of $\beta$ is essentially to change the number or rolls.
(b) Case of solitons. The effect of $\beta$ here is to change the pulse-width. }
\end{figure}
%%%%%%%%%%%%%%%%%%%%%%%%%%%%%%%%%%%%%%%%%%%%%%%%%%%%%%%%%%%%%%%%%%%%%%%%%%%%%%%%%%%%%%%%%%%%%%%%%%%%%%%%%%%%%%%%%%%%%%%%%%%%%%%%%%%%%%%%%%%%%%%%%%%%%%%%%%%
%%%%%%%%%%%%%%%%%%%%%%%%%%%%%%%%%%%%%%%%%%%%%%%%%%%%%%%%%%%%%%%%%%%%%%%%%%%%%%%%%%%%%%%%%%%%%%%%%%%%%%%%%%%%%%%%%%%%%%%%%%%%%%%%%%%%%%%%%%%%%%%%%%%%%%%%%%%
%%%%%%%%%%%%%%%%%%%%%%%%%%%%%%%%%%%%%%%%%%%%%%%%%%%%%%%%%%%%%%%%%%%%%%%%%%%%%%%%%%%%%%%%%%%%%%%%%%%%%%%%%%%%%%%%%%%%%%%%%%%%%%%%%%%%%%%%%%%%%%%%%%%%%%%%%%%

\section{Chaos}
\label{Chaos}

It is well known that chaos can potentially arise in any nonlinear system with at least three degrees of freedom.
The LLE is indeed a highly nonlinear and infinite-dimensional system, and in fact, the phenomenology of interest (Kerr comb generation) for us essentially relies on the nonlinearity. 

From a practical viewpoint, almost all high-dimensional and nonlinear systems display chaos when they are strongly excited. 
Chaos in Kerr combs has been unambiguously spotted both theoretically and experimentally in ref.~\cite{YanneNanPRL}, where the Lyapunov exponent had been computed and shown to be positive under certain circumstances. 
Another study on this topic is ref.~\cite{Matsko_Chaos}.

From the preceding sections, at least two routes to chaos can be identified in this system.

The first route corresponds to unstable Turing patterns.
In Figs.~\ref{chaos}(a) and~(b) shows that in that case, the Kerr comb is made of very strong spectral lines corresponding to the primary comb, and apparent spectral lines standing in between, and which are the signature to what was referred to as the secondary comb in refs.~\cite{YanneNanPRL,YanneNanPRA}. Hence, in this case, the route to chaos as the pump power $F^2$ is increased is a sequence of bifurcations starting with the primary comb which becomes unstable and leads to the emergence of a secondary comb; later on, higher-order combs are sequentially generated until a fully developed chaotic state is reached, as it can be seen in Figs.~\ref{chaos}(c) and~(d). 

The second route to chaos corresponds to unstable solitons.
Here, as the pump power is increased, the solitons become unstable and the system enters into a ``turbulent'' regime
characterized by the pseudo-random emergence of sharp and powerful peaks, as it can be seen in Figs.~\ref{chaos}(e).
This kind of chaos is very likely give birth in the WGM resonator to the extreme events referred to as \textit{rogue waves}.

\section{Influence of the dispersion parameter $\beta$}
\label{influencedispersion}

The bifurcation map displayed in Fig.~\ref{Figurebifnotatscale} in the  $\alpha$--$F$ plane disregards the effect of the magnitude of the dispersion. 
Indeed, the effect induced by the value of $|\beta|$ depends on the localized or non-localized nature of solution under study.

More precisely, if we consider both super- and sub-critical Turing patterns (non-localized structures) the effect of decreasing $|\beta|$ is straightforward as it increases the number of rolls according to Eq.~(\ref{def_l_mgm}) just above the pump. However, as the pump is increased beyond the bifurcation, such a decrease of $|\beta|$  also reduces the pulse-width of the individual Turing rolls in the sub-critical case. 
As far as solitons are concerned (localized structures), the effect of reducing  $|\beta|$ is essentially to decrease the pulse width of the solitons. Sub-critical Turing patterns and solitons have the same behavior in this regards, and this is a direct consequence of the fact that they are intimately connected from a topological point of view.
It should be noted once again that reducing the magnitude of second order GVD to arbitrarily small values can increases the relevance of higher-order dispersion terms in the Lugiato-Lefever model~\cite{PRA_Yanne-Curtis}. 

A general consequence of a decrease in  $|\beta|$ is that as the patterns have a narrower pulse-width, and as we commented in Part~I~\cite{Part_I}, this will generally allow for the excitation of a large number of solitons in the cavity.
Accordingly, the corresponding Kerr comb spectra will also display a higher complexity.\\

\section{Conclusion}
\label{Conclusion}

In this article, we have investigated the bifurcation structure
of the Lugiato-Lefever equation which is used to model Kerr comb
generation using WGM resonators.
We have focused on the case of anomalous dispersion and our analysis
has evidenced a plethora of possible steady states.
Turing patterns arise in the system through a $(i \omega)^2$ bifurcation,
and they can be characterized by their super- or sub-critical nature.
The threshold for these patterns can be analytically determined,
as well as the number of rolls. We have also been able to determine analytically 
the parametric gain allowing modulational instability.
We have also shown how the super- or sub-critical nature of the patterns affect their 
temporal dynamics and the way they can be excited.
Our investigations have also enabled to analyze the formation of 
solitons. We have shown that more complex structures, referred to a soliton molecules, can 
be generated as well in the system, and can eventually coexist along the azimuthal direction
of the resonator. Breather solitons have been analyzed as well, and we have also investigated
the emergence of chaos in the system.
The complexity of the related Kerr combs has also been studied and discussed.
Future work will be devoted to the investigation of the effect of higher-order nonlinearity and
dispersion~\cite{PRA_Yanne-Curtis,Pere_I,Pere_II}, 
and to the tailoring of the spectral characteristics of the combs for various technological applications~\cite{YanneNanOL,Engeneered_comb,Gaeta_LLE}.

\section*{Acknowledgements}

The authors acknowledge financial support from
the European Research Council through the project
NextPhase (ERC StG 278616). They would also like to
thank Mariana Haragus for very insightful comments and
suggestions. 
Authors would also like to acknowledge for the support
of the \textit{M\'esocentre de Calcul de Franche-Comt\'e}.

\end{document}